\documentclass[fleqn,usenatbib]{mnras}

\usepackage[T1]{fontenc}
\DeclareRobustCommand{\VAN}[3]{#2}
\let\VANthebibliography\thebibliography
\def\thebibliography{\DeclareRobustCommand{\VAN}[3]{##3}\VANthebibliography}

\usepackage{graphicx}	% Including figure files
\usepackage{amsmath}	% Advanced maths commands
\usepackage{amssymb}	% Extra maths symbols
\usepackage{subcaption}
\usepackage{tikz}
\usepackage{scalerel}
\usepackage{comment}
\usetikzlibrary{svg.path}
\usepackage{newtxtext,newtxmath}
\definecolor{orcidlogocol}{HTML}{A6CE39}
\tikzset{
  orcidlogo/.pic={
    \fill[orcidlogocol] svg{M256,128c0,70.7-57.3,128-128,128C57.3,256,0,198.7,0,128C0,57.3,57.3,0,128,0C198.7,0,256,57.3,256,128z};
    \fill[white] svg{M86.3,186.2H70.9V79.1h15.4v48.4V186.2z}
                 svg{M108.9,79.1h41.6c39.6,0,57,28.3,57,53.6c0,27.5-21.5,53.6-56.8,53.6h-41.8V79.1z M124.3,172.4h24.5c34.9,0,42.9-26.5,42.9-39.7c0-21.5-13.7-39.7-43.7-39.7h-23.7V172.4z}
                 svg{M88.7,56.8c0,5.5-4.5,10.1-10.1,10.1c-5.6,0-10.1-4.6-10.1-10.1c0-5.6,4.5-10.1,10.1-10.1C84.2,46.7,88.7,51.3,88.7,56.8z};
  }
}

\newcommand\orcidicon[1]{\href{https://orcid.org/#1}{\mbox{\scalerel*{
\begin{tikzpicture}[yscale=-1,transform shape]
\pic{orcidlogo};
\end{tikzpicture}
}{|}}}}

\newcommand{\aref}[1]{\hyperref[#1]{Appendix~\ref{#1}}}
\newcommand{\vai}{{v_{\text{A,ion}}}}

\newcommand{\vao}{{v_{\text{A0}}}}
\newcommand{\va}{v_{\text{A}}}
\newcommand{\Bo}{\mathbf{B}_{0}}
\newcommand{\Ma}{\mathcal{M}_{\text{A}}}
\newcommand{\Mao}{\mathcal{M}_{\text{A0}}}
\newcommand{\M}{\mathcal{M}}
\newcommand{\Exp}[1]{\left\langle #1 \right\rangle}
\newcommand{\vstr}{v_{\rm str}}
\renewcommand{\d}[1]{\ensuremath{\operatorname{d}\!{#1}}}

\title[Cosmic ray diffusion]{Turbulent diffusion of streaming cosmic rays in compressible, partially ionised plasma}

\author[Sampson, et al., 2022]{
Matt L. Sampson,$^{\orcidicon{0000-0001-5748-5393}\,1}$\thanks{E-mail: matthew.sampson@anu.edu.au}
James R. Beattie$^{\orcidicon{0000-0001-9199-7771}\,1}$,
Mark R. Krumholz$^{\orcidicon{0000-0003-3893-854X}\,1,2}$,
Roland M. Crocker$^{\orcidicon{0000-0002-2036-2426}\,1}$,
\newauthor
\ Christoph Federrath$^{\orcidicon{0000-0002-0706-2306}\,1,2}$ and Amit Seta$^{\orcidicon{0000-0001-9708-0286}\,1}$
\\
$^{1}$Research School of Astronomy and Astrophysics, Australian National University, Canberra, ACT 2611, Australia \\
$^{2}$Australian Research Council Centre of Excellence in All Sky Astrophysics (ASTRO3D), Canberra, ACT 2611, Australia 
}

\date{Accepted XXX. Received YYY; in original form ZZZ}

\pubyear{2022}

\begin{document}
\label{firstpage}
\pagerange{\pageref{firstpage}--\pageref{lastpage}}
\maketitle

\begin{abstract}
Cosmic rays (CRs) are a dynamically important component of the interstellar medium (ISM) of galaxies. The $\sim$GeV CRs that carry most CR energy and pressure are likely confined by self-generated turbulence, leading them to stream along magnetic field lines at the ion Alfv\'en speed. However, the consequences of self-confinement for CR propagation on 
%galactic 
galaxy
scales remain highly uncertain. In this paper, we use a large ensemble of magnetohydrodynamical turbulence simulations to quantify how the basic parameters describing ISM turbulence -- the sonic Mach number, $\M$ (plasma compressibility), Alfv\'en Mach number, $\Mao$ (strength of the large-scale field with respect to the turbulence), and ionisation fraction by mass, $\chi$ -- affect the transport of streaming CRs. We show that the large-scale transport of CRs whose small-scale motion consists of streaming along field lines is well described as a combination of streaming along the mean field and superdiffusion both along (parallel to) and across (perpendicular to) it; $\Mao$ drives the level of anisotropy between parallel and perpendicular diffusion and $\chi$ modulates the magnitude of the diffusion coefficients, while in our choice of units, $\mathcal{M}$ is unimportant except in the sub-Alfv\'enic ($\Mao \lesssim 0.5$) regime. Our finding that superdiffusion is ubiquitous potentially explains the apparent discrepancy between CR diffusion coefficients inferred from measurements close to individual sources compared to those measured on larger, Galactic scales. Finally, we present empirical fits for the diffusion coefficients as a function of plasma parameters that may be used as sub-grid recipes for global interstellar medium, galaxy or cosmological simulations.
\end{abstract}


%%%%%%%%%%%%%%%%%%%%%%%%%%%%%%%%%%%%%%%%%%%%%%%%%%

%%%%%%%%%%%%%%%%% BODY OF PAPER %%%%%%%%%%%%%%%%%%

\section{Introduction}
\label{sec:intro}
The role that the %relativistic 
non-thermal
particles known as cosmic rays (CRs) play in both star formation and galaxy evolution is one of the largest open questions in modern astronomy. Within the diffuse interstellar medium (ISM), CRs are important dynamically because their energy densities -- directly measured 
at the Earth and indirectly inferred at larger distances 
in the Milky Way and  in extragalactic systems -- are comparable to those in other interstellar reservoirs, such as turbulent motions of gas, magnetic fields, interstellar radiation, and self-gravity \citep{spitzer1978physical,boulares1990galactic,ferriere2001interstellar,draine2010physics,Grenier2015}. As a result, CRs may play an important role in either initiating or sustaining galactic winds and in regulating star formation \citep[e.g.,][]{socrates2008eddington, salem2014cosmological, simpson2016role, pakmor2016galactic, 2016ApJ...816L..19G,2017ApJ...834..208R, mao2018galactic, hopkins2021cosmic, hopkins2021effects, crocker2021cosmicb, crocker2021cosmic}. In the denser parts of the ISM that are shielded from ultraviolet starlight, CRs play a vital role in determining the distribution of thermal energy and the ionisation state and in initiating many of the chemical reaction chains that give rise to the formation of molecules \citep{cesarsky1978cosmic, everett2011interaction, padovani2009cosmic,glover2010modelling,grassi2014krome, drury2012turbulent, padovani2020impact}.

%However, the actual role  of CRs is uncertain due to the complex nature of their propagation and interaction with other components of the ISM. 

CRs having comparable energy density to other components of the ISM is a necessary but not sufficient condition for them to be dynamically important. To be dynamically important, sufficiently large gradients in CR energy densities must develop; as the transport of CRs largely determines these gradients, it is important to get an accurate description of the diffusion of CRs.

Since they are charged, CRs propagate via spiralling around magnetic field-lines under the Lorentz force. This motion means CRs are coupled to the magnetic fields existing in astrophysical plasmas \citep{strong2007cosmic, grenier2015nine, Grenier2015, gabici2019origin}, but it also means that CRs can interact resonantly with Alfv\'en waves whose wavelengths are comparable to their radius of gyration. These resonant interactions scatter CRs by altering the pitch angle between the CR velocity vector and the local magnetic field. The Alfv\'en waves responsible for resonant scattering can either be part of a turbulent cascade initiated by large-scale disturbances of the ISM (e.g., supernova blast waves), or can be generated by the CRs themselves via the streaming instability \citep[e.g.,][]{Kulsrud69a, Wentzel1974, Farmer04a, bell2013cosmic}. Low energy CRs ($\lesssim 10-100$ GeV in Milky Way type galaxies and up to $\sim 10$ TeV in starbursts -- \citealt{Krumholz2020CosmicGalaxies}), which dominate the total CR energy budget, are numerous enough and have small enough gyroradii that self-excitation is likely dominant for them. This leads to a situation of self-confinement whereby CRs stream along magnetic field lines at the same speed as the Alfv\'en waves they generate in the ionised component of the gas, and in a direction opposite the gradient in CR pressure\footnotemark \footnotetext{It should be made clear here that individual CR particles still move at velocity $\sim$ speed of light. However, due to the constant randomisation in pitch angle, the \textit{mean} position of self confined CR populations propagates along the magnetic field lines at roughly the ion Alfv\'en speed.}. In this paper, we will refer to populations of CRs that have been self-confined to stream along magnetic field lines at roughly the ionic Alfv\'en speed as streaming cosmic rays (SCRs). 

%The effects of these interactions on larger scale (ie. outer scale of turbulence) CR propagation are not well studied. 

While streaming may be the correct description of CR transport on scales comparable to CR gyroradii ($\ll 1 \ \rm{pc}$ for $\sim$GeV CRs), there is ample evidence that CR transport can be described approximately as diffusive when measured on molecular cloud or even galactic scales \citep[e.g.,][]{Krumholz2020CosmicGalaxies}. Direct \textit{in situ} observations of high-energy CRs reaching the Solar system indicate that their directions of travel are very close to isotropic, as would be expected for diffusive transport; models based on diffusion have successfully reproduced a large number of observations \citep[e.g.,][and references therein]{strong2007cosmic, zweibel2017basis}. Nor is it surprising that such a description applies: even if CRs move purely by streaming along field lines, interstellar plasmas are turbulent \citep[see][for a review]{federrath2016magnetic}, and so the field lines themselves are neither straight nor time independent. Thus, even if on small scales CRs did not diffuse at all, just the turbulent motion of the magnetic field lines to which they are bound should induce diffusion-like behaviour.

In principle, it should be possible to compute diffusion coefficients to describe this process in terms of the parameters that describe the magnetised turbulence, most prominently the sonic Mach number $\M$, Alfv\'en Mach number $\mathcal{M}_A$, and ionisation fraction $\chi$. These diffusion coefficients could be quite different from the traditional spatial diffusion coefficient for CRs that can be computed from, e.g., quasi-linear theory (QLT), because they describe a very different process, averaged over very different scales. The traditional CR diffusion coefficient from QLT is fundamentally a result of CRs making random walks in pitch angle, which leads them to perform a corresponding random walk in space along the magnetic field. The characteristic size scale over which it is reasonable to describe this process as a random walk, and thus as diffusive, is the mean distance that CRs travel before resonant scattering off incoherent Alfv\'en waves causes their pitch angles to randomise; this is much smaller than the characteristic scales associated with interstellar turbulence. By contrast, the diffusion coefficients associated with turbulent motion of the gas can only be defined on scales comparable to the sizes of turbulent eddies in the flow, and the diffusion they describe does not depend on rates of pitch angle scattering. Indeed, the turbulent diffusion could be non-zero even if the CR pitch angle distribution were a $\delta$-function, corresponding to no small-scale diffusion at all. Similarly, even though random walks in pitch angle only induce diffusion along field lines (at least to linear order), turbulence can induce diffusion perpendicular to the direction of the mean field.

Since most observational constraints on the rate at which CRs diffuse are sensitive to scales comparable to (or greater than) the scales of interstellar turbulence, rather than the much smaller scales of CR isotropisation, the effective rate of diffusion expected for SCRs at those larger scales is of considerable interest. While there have been efforts to develop theories for these quantities \citep{yan2002scattering, yan2004cosmic, shalchi2004nonlinear, shalchi2009analytical, lazarian2006cosmic, yan2008cosmic, beresnyak2011numerical, zweibel2013microphysics, evoli2014cosmic,cohet2016cosmic, Shukurov2017, zhao2017cosmic, Krumholz2020CosmicGalaxies, dundovic2020novel,reichherzer2020turbulence, reichherzer2021regimes}, robust parameter studies probing the relations between the plasma properties and CR diffusion over a wide range of parameter space are lacking. Thus, we still lack a complete effective theory of CR transport that can be used in simulations or models that do not resolve the characteristic scales on which CR transport becomes effectively diffusive, and thus a pure streaming description becomes inadequate. 

This issue is of broad importance for understanding the role of CRs in the ISM. Thus far attempts to address it have mostly proceeded empirically, for example by carrying out simulations or making models using a wide range of candidate CR transport prescriptions and seeing which ones best match observations \citep[e.g.][]{gabici2010constraints, Johannesson16a, Lopez18a, Chan19a, hopkins2021effects, crocker2021cosmic}. However, this approach has its limitations: it can determine what parameter values best fit the data within the context of a particular model but cannot tell us if that model is missing essential ingredients. Nor do these observations, which are largely limited to the Milky Way and its nearest neighbours, provide much insight into how CR transport might differ in more distant galaxies whose interstellar environments differ from those found locally \citep[e.g.,][]{Krumholz2020CosmicGalaxies}. In this paper, we therefore attempt a different approach: assuming that self-confinement and streaming are the most relevant mechanisms for CR transport on small scales, at least for the low-energy CRs that dominate the CR pressure budget, we seek to determine an effective theory for CR diffusion when measured over larger scales. We do so over a very wide range of plasma parameters, combining numerical simulations of MHD turbulence with those of CR transport through this turbulence and using the results as a series of numerical experiments to which we can fit a model.

The plan for the remainder of this paper is as follows. We describe the setup of our numerical experiments in \autoref{Methods} and show the results in \autoref{Results}. We use these results to build up an effective theory for CR transport in \autoref{Discussion} and summarise our findings in \autoref{Conclusions}.

\section{Methods}
\label{Methods}

Our goal is to measure the effective diffusion coefficient for CRs streaming along field lines in turbulent plasmas, across a wide range of plasma parameters. To this end, we construct a simulation and analysis pipeline in several steps. We first perform MHD simulations to produce background plasmas through which we can propagate CRs. We discuss the details of the MHD simulations in \autoref{sec:MHD_sim}. Our second step is to simulate the streaming of CRs through these plasmas, a process we describe in \autoref{sec:criptic}. In the final step, we construct a forward model for the CR position distribution that we can compare to our simulation results to infer large-scale diffusion parameters. We outline this model and our fitting method in \autoref{sec:diffuse_model}. 

%%%%%%%%%%%%%%%%%%%%%%%%%%%%%%%%%%%%%%%%%%%%%%%%%%%%%%%%%%%%%%%%%%%%%
%%% Trial parameters
%%%%%%%%%%%%%%%%%%%%%%%%%%%%%%%%%%%%%%%%%%%%%%%%%%%%%%%%%%%%%%%%%%%%%

We apply our pipeline to simulations at a range of sonic Mach number $\M$, Alfv\'en Mach number $\Mao$ and ionisation fraction (by mass) $\chi$. The first two of these describe the plasma itself, while $\chi$ affects the speed at which CRs stream, since the streaming speed is set by the ion Alfv\'en speed rather than the total Alfv\'en speed. We alter both $\M$ and $\Mao$ in the MHD simulation runs, while $\chi$ is an input parameter for the CR propagation simulation step. The parameter values we sample are $\M \in$ [2, 4, 6, 8, 10], $\Mao \in$ [0.1, 0.5, 1, 2, 4, 6, 8, 10] and $\log\chi \in$ [$-5,-4,-3,-2,-1,0$]. These values are chosen to capture the diversity of $\M$ and $\Ma$ encountered in the ISM \citep{tofflemire2011interstellar,burkhart2014measuring}. We carry out runs with every possible combination of these parameters, for a total of 240~trials. The naming convention for the trials is \texttt{M}$x$\texttt{MA}$y$\texttt{C}$z$ where $x$ is $\M$, $y$ is $\Mao$ (with the decimal point omitted) and $z$ is equal to $-\log\chi$. Thus, for example trial $\texttt{M2MA05C4}$ has $\M = 2$, $\Mao = 0.5$ and $\chi = 1 \times 10^{-4}$.

%%%%%%%%%%%%%%%%%%%%%%%%%%%%%%%%%%%%%%%%%%%%%%%%%%%%%%%%%%%%%%%%%%%%%
%%% Info on MHD simulations
%%%%%%%%%%%%%%%%%%%%%%%%%%%%%%%%%%%%%%%%%%%%%%%%%%%%%%%%%%%%%%%%%%%%%
\subsection{MHD simulations}
\label{sec:MHD_sim}
We generate turbulent MHD gas backgrounds through which we can propagate CRs using a modified version of the \textsc{flash} code \citep{fryxell2000flash,Dubey2008}, utilising a second-order conservative MUSCL-Hancock 5-wave approximate Riemann scheme \citep{bouchut2010multiwave,waagan2011robust,federrath2021sonic} to solve the 3D, ideal, isothermal MHD equations with a stochastic acceleration acting to drive the turbulence,
\begin{align} %%% u \to w
    \frac{\partial \rho}{\partial t} + \nabla\cdot(\rho \mathbf{w}) &= 0 \label{eq:continuity}, \\
    \rho\frac{\partial\mathbf{w}}{\partial t}  - \nabla\cdot \left[  \frac{1}{4\pi}\mathbf{B}\otimes\mathbf{B} -\rho\mathbf{w}\otimes\mathbf{w} - \left(c_s^2 \rho +\frac{B^2}{8\pi}\right)\mathbb{I}\right] &= \rho \mathbf{f}\label{eq:momentum}, \\
    \frac{\partial \mathbf{B}}{\partial t} - \nabla \times (\mathbf{w} \times \mathbf{B}) &= 0\label{eq:induction},\\
    \nabla \cdot \mathbf{B} &= 0, \label{eq:div0}
\end{align}
where  $\otimes$ is the tensor product, $\mathbb{I}$ is the identity matrix, $\mathbf{w}$ is the fluid velocity, $\rho$ the density, $\mathbf{B} = B_0 \mathbf{\hat{z}} + \delta\mathbf{B}(t)$ the magnetic field, with a constant mean (large-scale) field $B_0 \mathbf{\hat{z}}$, and turbulent field $\delta\mathbf{B}(t)$, $c_s$ is the sound speed, and $\mathbf{f}$ is the turbulent acceleration field. The simulation domain is a triply-periodic box of volume $\mathcal{V}=L^3$ and we drive the turbulence on the driving scale $\ell_0$ centred at $\ell_0 = L/2$. The time evolution of the driving field $\mathbf{f}$ follows an Ornstein-Uhlenbeck process with finite correlation time, 
\begin{align}\label{eq:turnover_time}
    \tau = \ell_0/\big<w^2\big>^{1/2}_{\mathcal{V}} = L/(2 c_s \M),
\end{align}
and is constructed such that we are able to set $2 \lesssim \M \lesssim 10$ and force with equal energy in both compressive $(\nabla\times\mathbf{f}=0)$ and solenoidal $(\nabla\cdot\mathbf{f}=0)$ modes. The driving is isotropic, and performed in $\mathbf{k}$ space, centred on $|\mathbf{k}L/2\pi|=2$ (corresponding to driving scale $\ell_0 = L/2)$ and falling off to zero with a parabolic spectrum within $1 \leq |\mathbf{k}L/2\pi| \leq 3$ (see \citealt{Federrath2008,Federrath2009,federrath2010comparing,federrath2022tg} for further details on the turbulence driving method).

We set the magnitude of the large-scale magnetic field component $B_0$ by specifying the desired Alfv\'en Mach number of the mean field, $\Mao$, the turbulent $\M$, and then requiring $B_0 = c_s \sqrt{4\pi\rho_0} \M / \Mao$. The initial velocity field is set to $\mathbf{w}(x,y,z,t=0)=\mathbf{0}$, with units $c_s=1$, the density field $\rho(x,y,z,t=0)=\rho_0$, with units $\rho_0=1$ and $\delta\mathbf{B}(t) = \mathbf{0}$, with units $c_s\rho_0^{1/2}$. For more details about the current simulations, we refer the reader to \citet{beattie2020magnetic}, \citet{Beattie2021_spdf}, and \citet{beattie2021multishock}.

We run the simulations for $10\,\tau$ but only use data from the last $5\,\tau$ because for $t < 5\tau$ the turbulence is not 
necessarily
fully developed for simulations with $\Mao < 1$ \citep{Beattie2021_spdf}, i.e., $\Exp{\mathbf{X}(t+\Delta t}_{\mathcal{V}} \neq \Exp{\mathbf{X}(t)}_{\mathcal{V}}$, for volume-averages of arbitrary field variable $\mathbf{X}$ and time interval $\Delta t$. However, we note for $\Mao \gtrsim 1$ the turbulence is fully developed within $2\,\tau$ \citep{federrath2010comparing}. We dump an MHD realisation of the 3D field variables every $t = \tau / 10$. For our main results, we discretise the $L^3$ domain into $576^3$ cells. Both the grid resolution and the temporal resolution are determined via convergence tests detailed in \aref{sec:converge}. The choice of resolution here is made based on the analysis of the CR propagation post-processing results, as opposed to an analysis of parameters of the MHD dataset itself. % note add more on convergence limitations here

\begin{figure*}
    \centering
    \includegraphics[width=0.87\linewidth]{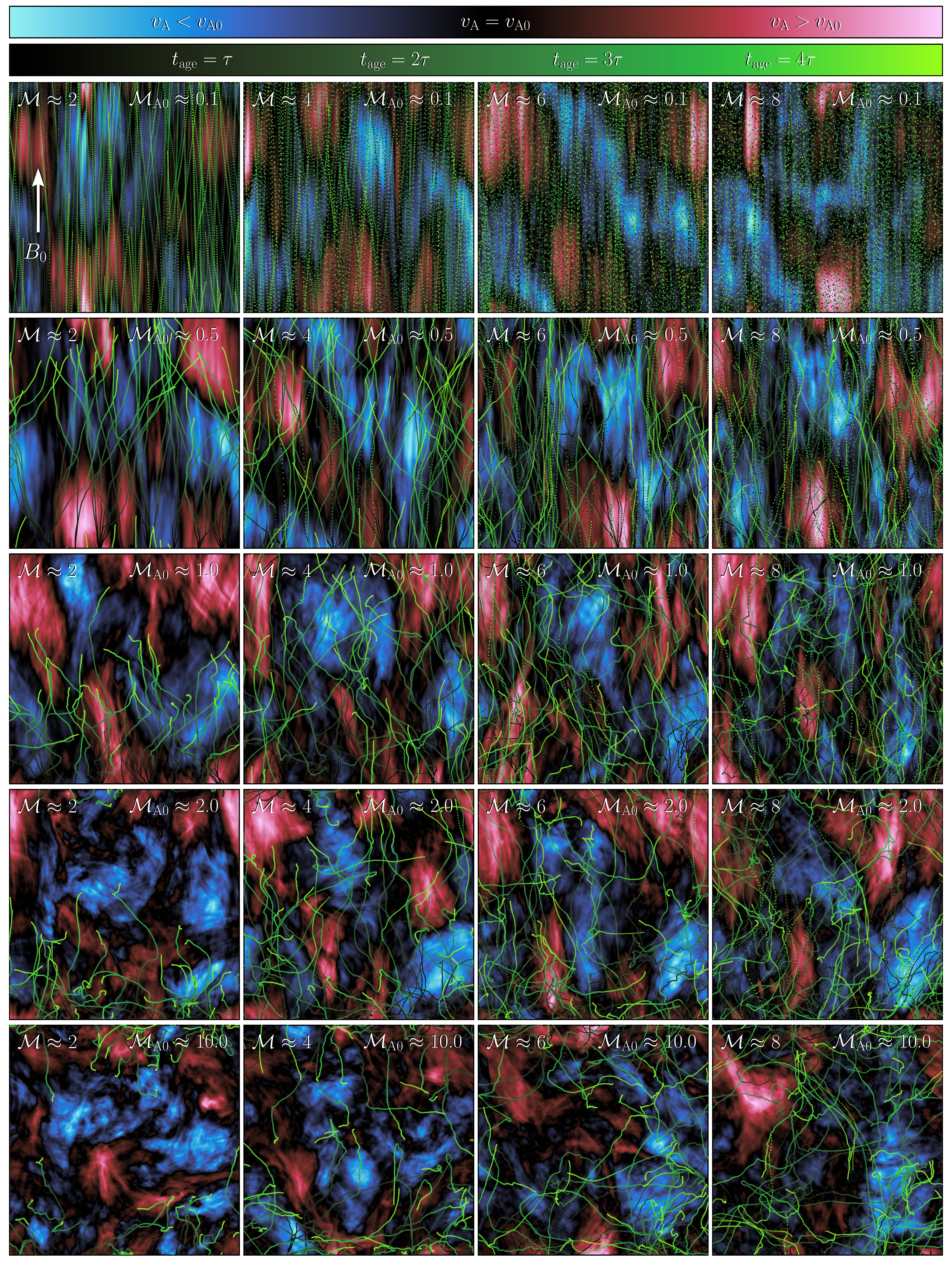}
    \caption{Alfv\'en velocity structure (background) and SCR packet positions (points) projected perpendicular to the direction of the mean magnetic field, $\Bo$. The field colour indicates the logarithmic streaming velocity along the projection direction, $\int \d{(\ell_{\perp}/L)} \, \va /\vao$, where $\vao = \Exp{\va}_{\mathcal{V}}$; red indicates locations where the streaming speed is above the mean, blue below the mean and black, equal to the mean. Points show the positions of sample CR packets through their time-evolution up until the age of the visualised gas background. Colour of the particles indicates their age, $t_{\rm age}$ in turbulent correlation times (black is early in the particles temporal evolution and green later). Simulations are organised by $\Mao$ (fixed $\Mao$ each row) and $\M$ (fixed $\M$ each column), where the simulations with the strongest magnetisation ($\Mao = 0.1$) and weakest compressibility ($\M = 2$) are shown in the top row, first column, and the weakest magnetisation ($\Mao = 10$) and strongest compressibility ($\M = 8$) in the bottom row, last column.}
    \label{fig:20_panel}
\end{figure*}

%%%%%%%%%%%%%%%%%%%%%%%%%%%%%%%%%%%%%%%%%%%%%%%%%%%%%%%%%%%%%%%%%%%%%
%%% CRIPTIC
%%%%%%%%%%%%%%%%%%%%%%%%%%%%%%%%%%%%%%%%%%%%%%%%%%%%%%%%%%%%%%%%%%%%%
\subsection{Cosmic ray propagation simulations}
\label{sec:criptic}

With the MHD simulations in hand, our next step is to simulate the propagation of CRs streaming through them. We envision that each of our simulation boxes is subject to a large-scale CR pressure gradient, on scales much larger than the box scale, so that CRs stream through the simulation box in a single direction. We further assume that our boxes are much larger than the size scale on which the CR pitch angle distribution isotropises or the scale over which the CR distribution function comes into equilibrium between growth of the streaming instability and damping of Alfv\'en waves. Under these assumptions, we can treat CR propagation within the simulations as simply streaming down field lines at the ion Alfv\'en speed, together with advection with the gas. Formally, we assume that the CR distribution function $f(\mathbf{x}, t)$ evolves as
\begin{equation} \label{eq:FPE}
    \frac{\partial f}{\partial t} = \nabla \cdot \left[\left(\mathbf{w} + v_{\rm str}\mathbf{b}\right) f \right]
    %\frac{\partial}{\partial x_i}\left[\left(w_i + v_{\rm str} b_i\right) f\right],
\end{equation}
where $\mathbf{w}$ is the gas bulk velocity, $v_{\rm str}$ is the streaming speed, and $\mathbf{b} = \mathbf{B}/|\mathbf{B}|$ is a unit vector parallel to the local magnetic field. The streaming speed in turn is equal to the ion Alfv\'en speed, and thus is a function of the local magnetic field strength $B$, density $\rho$, and the ionisation fraction $\chi$:\footnote{A corollary of this expression is that, if one adopts a CR streaming model where the streaming speed is not exactly equal to the ion Alfv\'en speed, but is instead some multiple of it, then this is completely equivalent to changing the value of $\chi$. That is, if one assumes CRs stream at 10 times the local ion Alfv\'en speed, then this is equivalent to simply reducing $\chi$ by a factor of $100$.}
\begin{equation}
    v_{\rm str} = \frac{B}{\sqrt{4\pi\chi\rho}} = \frac{1}{\Mao \sqrt{\chi}} \left(\frac{\ell_0}{\tau}\right) \left(\frac{B}{B_0}\right) \left(\frac{\rho}{\rho_0}\right)^{-1/2}.
\end{equation}

We solve \autoref{eq:FPE} using the CR propagation code \textsc{criptic}\footnote{ Available from \url{https://bitbucket.org/krumholz/criptic/src/master/}.} \citep{Krumholz2022}. \textsc{Criptic} solves the Fokker-Planck equation for the CR distribution function including gain and loss processes using a Monte Carlo approach whereby we follow the trajectories of sample CR packets. For the purposes of the simulations here, we disable all \textsc{criptic} functionality that describes diffusion or microphysical interactions between CRs and the gas, so the equation solved reduces to \autoref{eq:FPE}.

We initialise our \textsc{Criptic} simulations by placing a grid of $9\times 9$ sources in the plane perpendicular to $\Bo$, each of which injects CR sample packets into the simulation volume at a rate $\Gamma_{\rm inj} = n_{\rm inj}/\tau$, where we select $n_{\rm inj} = 10^6$ based on the convergence testing presented in \aref{sec:converge}. We evolve the injected CRs for $t=5\,\tau$ (starting at $t=5\tau$, so the turbulence has reached statistical steady state at the point where injection begins), and we use periodic boundary conditions on the CRs, consistent with the boundary conditions used in the MHD simulations.

\textsc{Criptic} needs to know the plasma state at arbitrary positions and times, so that it can evolve \autoref{eq:FPE}. To achieve this, we linearly interpolate the MHD simulation realisations (dumped at intervals of $\tau/10$) at every position, $\mathbf{x}_i$, and time, $t_i$ (noting $i$ subscript denotes for each individual CR packet). We use linear interpolation rather than a higher-order scheme because only linear interpolation maintains $\nabla\cdot\mathbf{B} = 0$. 

The output from each \textsc{criptic} simulation is a set of 3-dimensional CR positions dumped every $t = \tau / 10^3$. Each CR is labelled by the source from which it emerged and by the time at which it was injected. Hence, for each CR we know the current position, the starting position, and the amount of time for which the CR has been moving in the simulation.

\autoref{fig:20_panel} shows a visual example of the \textsc{Criptic} outputs. In this panel, CRs trajectories are shown in green, with age increasing from the darkest to lightest shading of green. Each panel represents a different MHD simulation with $\Mao$ increasing with the rows, and $\mathcal{M}$ increasing with the columns. We see in the low $\Mao$ trials (top 4 panels) the SCRs travel predominantly directly up the domain with little deviation from straight lines, while as $\Mao$ increases the trajectories become increasingly isotropic and random.

%%%%%%%%%%%%%%%%%%%%%%%%%%%%%%%%%%%%%%%%%%%%%%%%%%%%%%%%%%%%%%%%%%%%%
%%% Simulation trials
%%%%%%%%%%%%%%%%%%%%%%%%%%%%%%%%%%%%%%%%%%%%%%%%%%%%%%%%%%%%%%%%%%%%%

\subsection{CR diffusion model and fitting}
\label{sec:diffuse_model}
The third step in our simulation pipeline is to reduce the set of CR packet ages $t_i$ and displacements $x_i$ relative to the location of that packet's source (where $x$ here can mean the coordinate in either the direction parallel to the mean field, $z$, or perpendicular to it, $x$ or $y$ -- we carry out a separate fit for the parameters describing transport in each cardinal direction) output by our simulations to a set of summary statistics that allow us to compare macroscopic CR transport between trials with different plasma parameters. Here and through much of the rest of the paper, we work in a dimensionless unit system where the turbulent driving scale $\ell_0 = 1$ and the turbulent turnover time $\tau = 1$ (c.f.~\autoref{eq:turnover_time}); in this unit system, the streaming speed at the mean density and magnetic field in the simulation is $v_{\rm str,0} = 1/\sqrt{\chi}\Mao$.

Let $\boldsymbol{\theta}$ be a vector of parameters describing CR transport as a function of the plasma parameters $\mathcal{M}$, $\Mao$, and $\chi$; we wish to fit $\boldsymbol{\theta}$ from the set of CR packet positions and times $(x_i,t_i)$ determined from our simulations. From Bayes' theorem, the posterior probability density at a particular point in parameter space obeys
\begin{equation}
\label{eqn:bayes}
    p(\theta | \{x_i,t_i \}) = \mathcal{N} \mathcal{L}\left( \{x_i,t_i \} | \boldsymbol{\theta} \right) p_{\text{prior}}(\boldsymbol{\theta}),  
\end{equation}
where $p_{\text{prior}}(\boldsymbol{\theta})$ is the prior probability density, $\mathcal{N}$ is a normalisation factor chosen to ensure the integral over our probability density function $ = 1$, and $\mathcal{L}\left( \{x_i,t_i \} | \boldsymbol{\theta}\right)$ is the likelihood function evaluated at $\{x_i,t_i \}$ for a vector of parameters $\boldsymbol{\theta}$. 
With this formulation, we may use a Markov chain Monte Carlo (MCMC) fitting approach to generate the posterior distribution. We must now determine an appropriate likelihood function that specifies the probability density of the data ($x_i,t_i$) given $\boldsymbol{\theta}$.

For this study, we approximate CR transport to be well described by a linear combination of CRs streaming along magnetic field-lines and superdiffusive transport \citep{xu2013cosmic,lazarian2014superdiffusion,litvinenko2014analytical}. Often numerical studies on CR transport calculate diffusion coefficients directly from the second moment of the CR spatial distribution \citep{qin2009pitch,wang2019diffusion,xu2013cosmic,snodin2016global,Seta2018}. However, this method is only viable when spatial dispersion grows linearly with time, which is not the case if superdiffusion is present. We also expect to have a systematic drift of CRs in the $\mathbf{z}$ direction due to CR streaming, which motivates us to use a model adjusted to capture both superdiffusion and constant drift. 

\subsubsection{Likelihood function for  generalised diffusion}
To derive our model, we start from the simplest case of drift-free diffusion in an infinite domain, then add streaming and periodicity. Superdiffusive transport is characterised by a generalised diffusion equation, whereby the probability density $f(x,t)$ evolves as
\begin{equation}
\label{eqn:diffusion}
    \frac{\partial f}{\partial t} = \kappa\Delta^{\alpha}f,
\end{equation}
where $\Delta^{\alpha}$ is the fractional diffusion operator of order $\alpha$, where $\alpha \in \mathbb{R}$. Gaussian diffusion corresponds to $\alpha = 2$ in which case $\Delta^{\alpha} =\nabla^2$, the Laplacian operator. A general solution to \autoref{eqn:diffusion} may be written in terms of a Green's function that solves the initial value problem $f(x,0) = \delta(x)$, where $\delta$ is the Dirac delta function. For an infinite domain the Green's function is
\begin{equation}
\label{eqn:Green}
    G(x,t) = \mathcal{N} \int \exp(-ikx) \exp(-\kappa|k|^{\alpha}t)\,dk,
\end{equation}
where $\mathcal{N}$ is the normalisation factor to be chosen such that $\int G(x,t) \ dx = 1  \ \forall t$, and $\kappa$ is the generalised diffusion coefficient \citep{zaburdaev2015levy}. Note here that $\mathcal{N}$ may be an explicit function of time, such that $\mathcal{N} = \mathcal{N}(t)$. The integral in \autoref{eqn:Green} cannot be solved analytically for general $\alpha$, hence it is left represented in the Fourier domain. For the special case $\alpha = 2$, one can immediately see that \autoref{eqn:Green} reduces to the Fourier transform of a Gaussian, which is also a Gaussian.

For a CR packet of age $t_i$, the Green's function gives the PDF of position $x_i$, and thus the log likelihood function\footnote{Maximising the log likelihood function is equivalent to maximising the likelihood function  so we take the log likelihood for computational simplicity.} for our ensemble of CR packets is
\begin{equation}
\label{eqn:Levy}
      \text{ln} \ \mathcal{L}\left( \{x_i,t_i \} | \alpha,\kappa \right) \propto \sum_{i=1}^{N}   \text{ln} \ G(x_i,t_i | \alpha,\kappa),
\end{equation}
where the parameters are $\boldsymbol{\theta} = (\alpha, \kappa)$. Note that, at fixed age $t$, the spatial distribution $G(x,t)$ is a L\'evy stable distribution, which previous authors have found to be suitable for modelling CR and brownian-like diffusion \citep{zimbardo1995anomalous,lagutin2001fractional,liu2004numerical,litvinenko2014analytical,rocca2016general}. Our implementation uses the \textsc{pylevy} python package \citep{jose_maria_miotto_2016_53787}, which provides efficient numerical evaluation of integrals of the form given by \autoref{eqn:Green}.

To add streaming in the direction parallel to $\Bo$ (i.e., along $z$) to this picture, let $u$ be the mean streaming velocity, so that we replace our positional variable $x$ with $x_i - u t_i$. Since we fit in each direction independently, we have three streaming speeds, $u_x$, $u_y$, and $u_z$. We expect $u_x = u_y = 0$  due to symmetry, but we nonetheless keep them in our fitting pipeline as a check for sensible results. The inclusion of streaming transforms \autoref{eqn:Levy} into
\begin{equation}
\label{eqn:green_drift}
    \text{ln} \  \mathcal{L}\left( \{x_i,t_i \} | \alpha,\kappa, u \right) \propto \sum_i \text{ln} \ G(x_i - u t_i , t_i| \alpha,\kappa, u),
\end{equation}
where we now have three fit parameters, $\boldsymbol{\theta} = (\alpha, \kappa, u)$, and $u$ from here on is referred to as the drift parameter.

%%%%%%%%%%%%%%%%%%%%%%%%%%%%%%%%%%%%%%%%%%%%%%%%%%%%%%%%%%%%%%%%%%%%%
%%% Periodic bounds
%%%%%%%%%%%%%%%%%%%%%%%%%%%%%%%%%%%%%%%%%%%%%%%%%%%%%%%%%%%%%%%%%%%%%
\subsubsection{Periodic boundary conditions}
Thus far we have written down the likelihood function for an infinite domain. However, our simulations use periodic boundary conditions, which admit a different Green's function. To construct the periodic domain likelihood function we note that our periodic domain from $x=-1$ to $1$ (recalling that we work in units where the turbulent driving scale $\ell_0 = 1$, so the box length $L=2$) containing a single source at $x=0$ is completely identical to a non-periodic domain containing an infinite array of sources located at $x = nL$ for $n \in \left\{ -\infty,\hdots,-2,-1,0,1,2,\hdots,\infty \right\}$. We can therefore write the Green's function for a periodic domain of length $L$ as
\begin{equation}
\label{eqn:periodic}
    G_L(x,t) = \mathcal{N} \ \sum_{n = -\infty}^{\infty} G(x + nL,t),
\end{equation}
where $G(x,t)$ is our Green's function for the infinite domain. Here we note that $G(x,t)$ has the scaling behavior $G(x,t) \sim t / |x|^{4\alpha + 1}$. Since $\alpha > 0$ due to the definition of the fractional diffusion operator, \autoref{eqn:periodic} approaches a convergent geometric series for large $|n|$.

In practice, we approximate the infinite sum in \autoref{eqn:periodic} as follows. Let us define a finite value $N$ such that we may approximate $G_L(x,t)$ as 
\begin{equation}
\label{eqn:sum_green}
    G_L^{(N)}(x,t) \propto  \sum_{n = -N}^{N} G(x + nL,t),
\end{equation}
where the normalisation factor $\mathcal{N}$ has been omitted. Now we define an error estimate
\begin{equation}
    \varepsilon^{(N)}(x,t) = 1 - \frac{G_L^{(N-1)}(x,t)}{G_L^{(N)}(x,t)}.
\end{equation}
We can then approximate $G_L(x,t)$ by $G_L^{(N)}(x,t)$ evaluated with a value of $N$ chosen such that $\varepsilon^{(N)}(x,t) < \text{tol}$ for some specified tolerance parameter tol.
Thus, our final likelihood function is 
\begin{equation}
\label{eqn:logLike_final}
        \text{ln} \  \mathcal{L}\left( \{x_i,t_i \} | \alpha,\kappa, u \right) \propto \sum_i \text{ln} \ G_L(x_i - u t_i , t_i| \alpha,\kappa, u),
\end{equation}
where we approximate $G_L$ by $G_L^{(N)}$ evaluated with a tolerance $\text{tol} = 10^{-3}$.

\subsubsection{Fitting method and priors}
Now that we have written down the log likelihood function, \autoref{eqn:logLike_final}, we can compute the posterior probability, \autoref{eqn:bayes}, from Bayes' theorem. In practice, we carry out this calculation using \textsc{emcee} \citep{foreman2013emcee}. Following the conventions used in the \textsc{pylevy} \citep{pylevy}, instead of using $\alpha$, $u$, and $\kappa$ as our parameters, we instead use $\alpha$, $u$, and $c$, where $c = \kappa^{1/\alpha}$; however we report our results in terms of $\kappa$ rather than $c$.  We adopt flat priors over the range $0.75 < \alpha \leq 2$\footnote{We note here that an upper bound of $\alpha = 2$  is a requirement of the \textsc{pylevy} model. However, as $\alpha > 2$ implies sub-diffusion (which has not been suggested in the literature) we do not expect this limitation to have significant effects on our results.}, $c \geq 0$, and $u \geq 0 $. We fit each direction individually, i.e., we perform separate \textsc{emcee} evaluations for $x_i = x$, $y$, and $z$. For all trials, we use 48 walkers with 1000 iterations and a burn-in period of 400 steps. Visual inspection of the chains confirms that this burn-in period is sufficient for the posterior PDF to reach statistical steady state.

\section{Results}
\label{Results}

In this section, we summarise the outputs of our simulation and fitting pipeline. We begin in \autoref{ssec:example_results} by walking through one example case to illustrate the typical results of our fits, and then in the remainder of this section we summarise the broad trends that we observe in the drift speed $u$ (\autoref{sec:drift}), superdiffusion index $\alpha$ (\autoref{sec:alpha}), and diffusion coefficients $\kappa$ (\autoref{sec:kappa}) as we vary the simulation parameters. We report our fit results for all trials in \autoref{tab:all}.

\begin{table*}
\centering
\footnotesize
\caption{Results from MCMC fitting for sample of trials. Columns show, from left to right, the name of the trial, the generalised diffusion coefficients in the three cardinal directions, $\kappa_x,\kappa_y,\kappa_z$, the superdiffusion indices, $\alpha_x,\alpha_y,\alpha_z$, and the drift parameters, $u_x,u_y,u_z$. For each quantity we report the $50^{\rm{th}}$ percentile of the marginal posterior PDF for that parameter, with the superscript denoting the $84^{\rm{th}} - 50^{\rm{th}}$ percentile and the subscript the $50^{\rm{th}} - 16^{\rm{th}}$ percentile ranges. All parameters expressed in a unit system whereby positions are measured in units of the turbulent driving length $\ell_0$, and times in units of the turbulent turnover time $\tau$. This table is a stub to illustrate form and content.The full table is available in the electronic version of this paper and at CDS 
via \texttt{https://cdsarc.unistra.fr/viz-bin/cat/J/MNRAS.}}
\rotatebox{0}{%
\label{tab:all}
\begin{tabular}{llllllllll} \hline \hline
Trial & $\kappa_x$ & $\kappa_y$ & $\kappa_z$ &  $\alpha_x$ & $\alpha_y$ & $\alpha_z$ & $u_x$ & $u_y$ & $u_z$    \\\hline \hline \\
\texttt{M2MA01C0}  & $0.039_{0.003}^{0.002} $ & $0.027_{0.002}^{0.002} $ & $0.239_{0.004}^{0.005}$ & $1.450_{0.010}^{0.007}$ & $1.509_{0.010}^{0.007}$ & $1.538_{0.010}^{0.010}$ & $0.032_{0.004}^{0.003}$ & $0.000_{0.000}^{0.020}$ & $10.61_{0.016}^{0.020}$ \\ \\
\texttt{M2MA01C1}  & $0.051_{0.001}^{0.000} $ & $0.035_{0.001}^{0.000} $ & $0.211_{0.007}^{0.004} $ & $1.449_{0.012}^{0.002} $ & $1.510_{0.010}^{0.002} $ & $1.883_{0.015}^{0.011} $ & $0.033_{0.004}^{0.004} $ & $0.000_{0.000}^{0.02} $  & $32.69_{0.030}^{0.030} $  \\ \\
\texttt{M2MA01C2}  & $0.038_{0.002}^{0.001} $ & $0.26_{0.001}^{0.001} $ & $0.459_{0.010}^{0.009} $ & $1.468_{0.011}^{0.016} $ & $1.521_{0.014}^{0.016} $ & $1.813_{0.008}^{0.008} $ & $0.032_{0.004}^{0.004} $ & $0.000_{0.000}^{0.02} $  & $102.7_{0.040}^{0.048} $ \\ \\
$\cdot$ & $\cdot$ & $\cdot$ & $\cdot$ & $\cdot$ & $\cdot$ & $\cdot$ & $\cdot$ & $\cdot$ & $\cdot$ \\
$\cdot$ & $\cdot$ & $\cdot$ & $\cdot$ & $\cdot$ & $\cdot$ & $\cdot$ & $\cdot$ & $\cdot$ & $\cdot$ \\
$\cdot$ & $\cdot$ & $\cdot$ & $\cdot$ & $\cdot$ & $\cdot$ & $\cdot$ & $\cdot$ & $\cdot$ & $\cdot$ \\ 
\texttt{M10MA10C5}  & $2.242_{0.110}^{0.142} $ & $2.025_{0.102}^{0.209} $ & $2.659_{0.041}^{0.042} $ & $1.673_{0.015}^{0.013} $ & $1.657_{0.003}^{0.010} $ & $1.604_{0.015}^{0.018} $ & $0.229_{0.034}^{0.043} $ & $0.165_{0.002}^{1.031} $  & $3.042_{0.0.041}^{0.038} $ \\ \\ \hline \hline
\end{tabular}%
}
\end{table*}

\subsection{Example case}
%\label{sec:mcmc}
\label{ssec:example_results}

\begin{figure*}
    \centering
    \includegraphics[width=\linewidth]{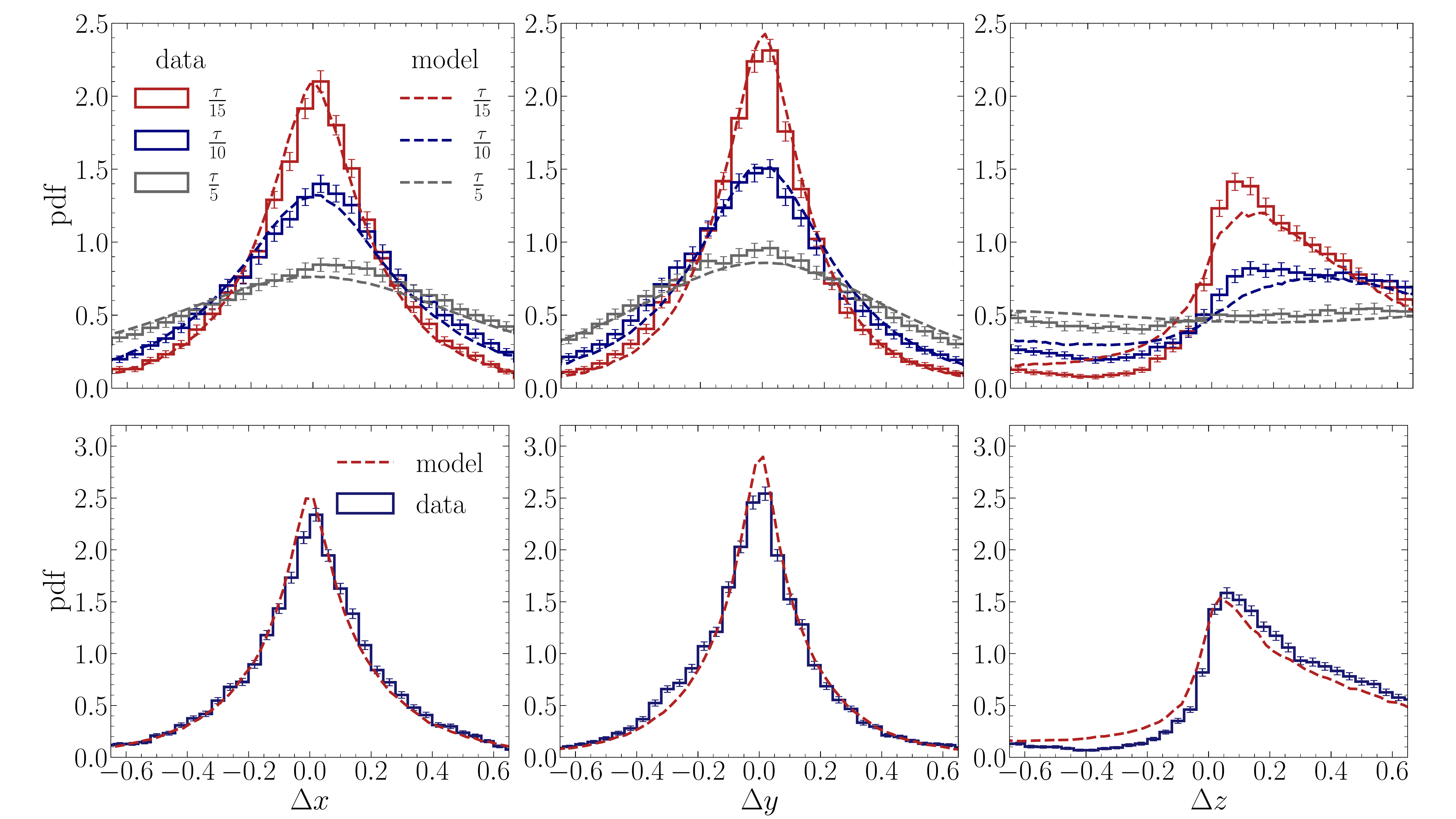}
    \caption{Probability distribution of SCR displacements $\Delta x$, $\Delta y$, $\Delta z$ in run \texttt{M6MA4C3} for SCR age slices of $\sim$ $\tau / 15$, $\tau / 10$, and $\tau / 5$ (\textbf{upper panels}), and for an integrated distribution of SCRs of age $<\tau / 5$ (\textbf{lower panels}); positions are measured normalised to the turbulent driving scale $\ell_0$, so the periodic simulation box goes from $-1$ to $1$. Histograms indicate simulation results, with error bars showing the Poisson uncertainty from the finite number of SCRs in each bin. Dashed lines show our streaming plus superdiffusion transport model, evaluated using the $50^{\rm{th}}$ percentile values of the fit parameters $\kappa$, $u$, and $\alpha$.}
    \label{fig:fits}
\end{figure*}

We begin by examining in detail the results for trial \texttt{M6MA4C4} ($\M = 6, \Mao = 4$, $\chi = 10^{-4}$), in order to illustrate the nature of the results and the action of our fitting pipeline. In \autoref{fig:fits}, we show histograms of the SCR displacements in the three cardinal directions, $\Delta x$, $\Delta y$, and $\Delta z$, at the final output time, $10\tau$. The upper panels show SCR distributions in narrow age windows $t = (0.25\pm 0.025)\tau$ (i.e., CRs with an age $ t \sim \tau / 15$), $\tau / 10$, and $\tau / 5$, while the bottom panels show the distribution for all SCRs with ages $<\tau / 5$. As expected, the CR displacement distributions in the $x$ and $y$ directions are symmetric about zero, with the width of the distribution increasing with time. In the $z$ direction, parallel to the large-scale magnetic field, streaming induces a substantial asymmetry, so that, particularly at young ages, more SCR packets have $\Delta z>0$ (i.e., in the direction of streaming along the mean field) than $\Delta z < 0$.

For comparison, the dashed lines in the figure show the predicted distribution of SCR positions at the corresponding ages (integrated over age for the lower panels) for our superdiffusion plus streaming model, evaluated using the $50^{\rm th}$ percentile values of the fit parameters $\kappa$, $u$, and $\alpha$. We can see the model matches the data very closely; while there is a slight systematic overestimation of the central bins of the integrated data, the shapes of the wings and the asymmetry in the parallel direction due to streaming are well-captured.

\begin{figure}
\centering
\includegraphics[width=\linewidth]{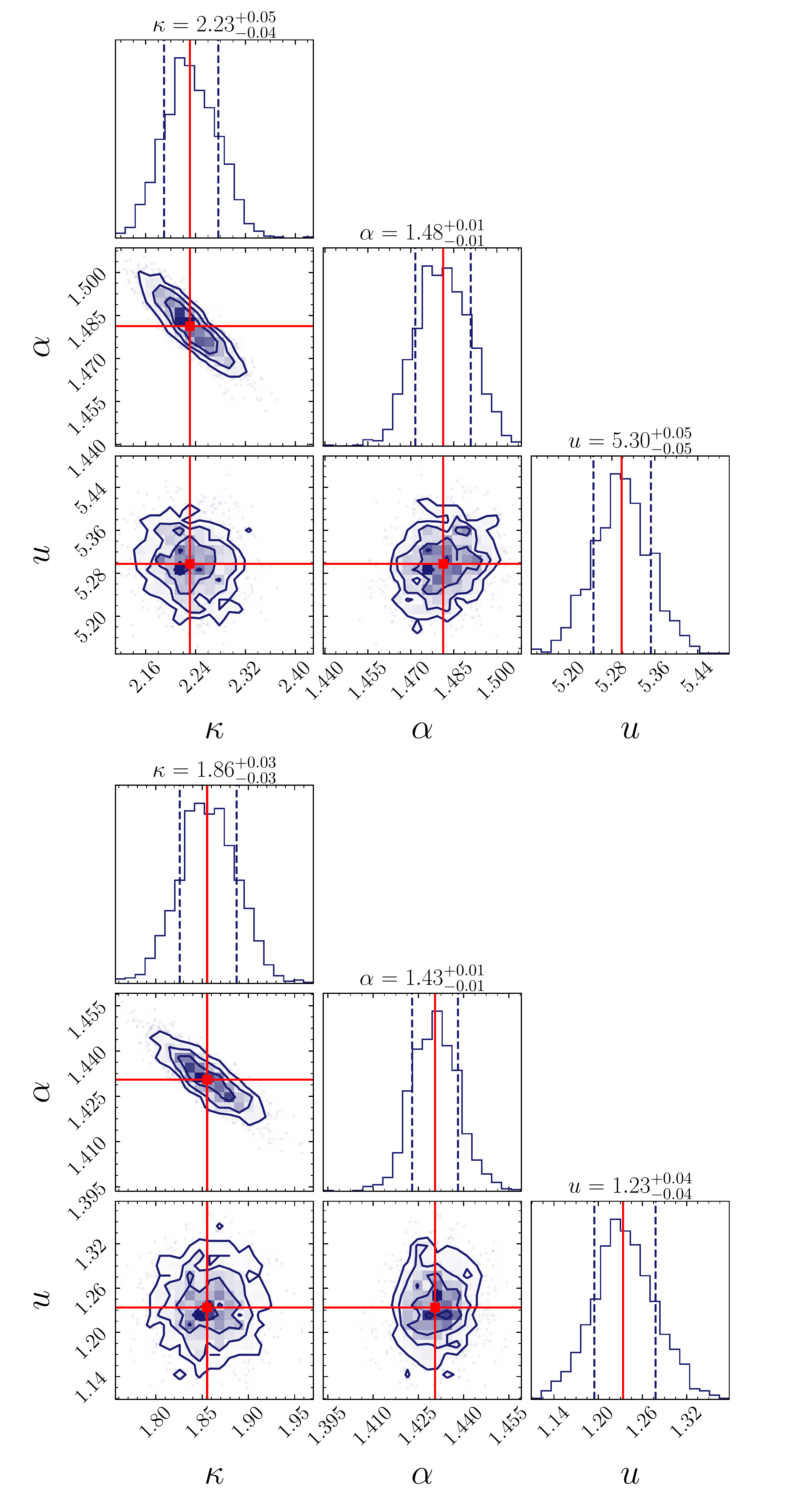} 
\caption{Corner plots showing posterior distributions from MCMC fitting for the three parameters $\kappa, \alpha$, and $u$ for trial \texttt{M6MA4C4}. At the top of each column, we report the $16^{\rm{th}}$, $50^{\rm{th}}$, and $84^{\rm{th}}$ percentile values; $50^{\rm{th}}$ percentile values are also indicated by the red lines in the plots, while in the histograms, dashed vertical lines show the $16^{\rm{th}}$ to $84^{\rm{th}}$ percentile range. \textbf{Upper panel:}  We show results for the $x$ direction (perpendicular to $\Bo$). \textbf{Lower panel:} shows the same results for the $z$ direction (parallel to $\Bo$). In both cases we see very small uncertainties (range between $16^{\rm{th}}$ and $84^{\rm{th}}$ percentiles) on the $50^{\rm{th}}$ percentile. Note we omit the corner plot for $y$ due to its similarity to $x$.}
\label{fig:MCMC}
\end{figure}

\autoref{fig:MCMC} shows the posterior distributions for the fit parameters we obtain for this case; the upper panel shows one of the two directions perpendicular to the mean magnetic field direction, while the lower panel shows the direction along the mean field. We see that in all cases the fit parameters are tightly constrained with small uncertainties.

\subsection{Drift speed: \texorpdfstring{$u$}{Lg}}
\label{sec:drift}
\begin{figure*}
\centering
\includegraphics[width=0.95\linewidth]{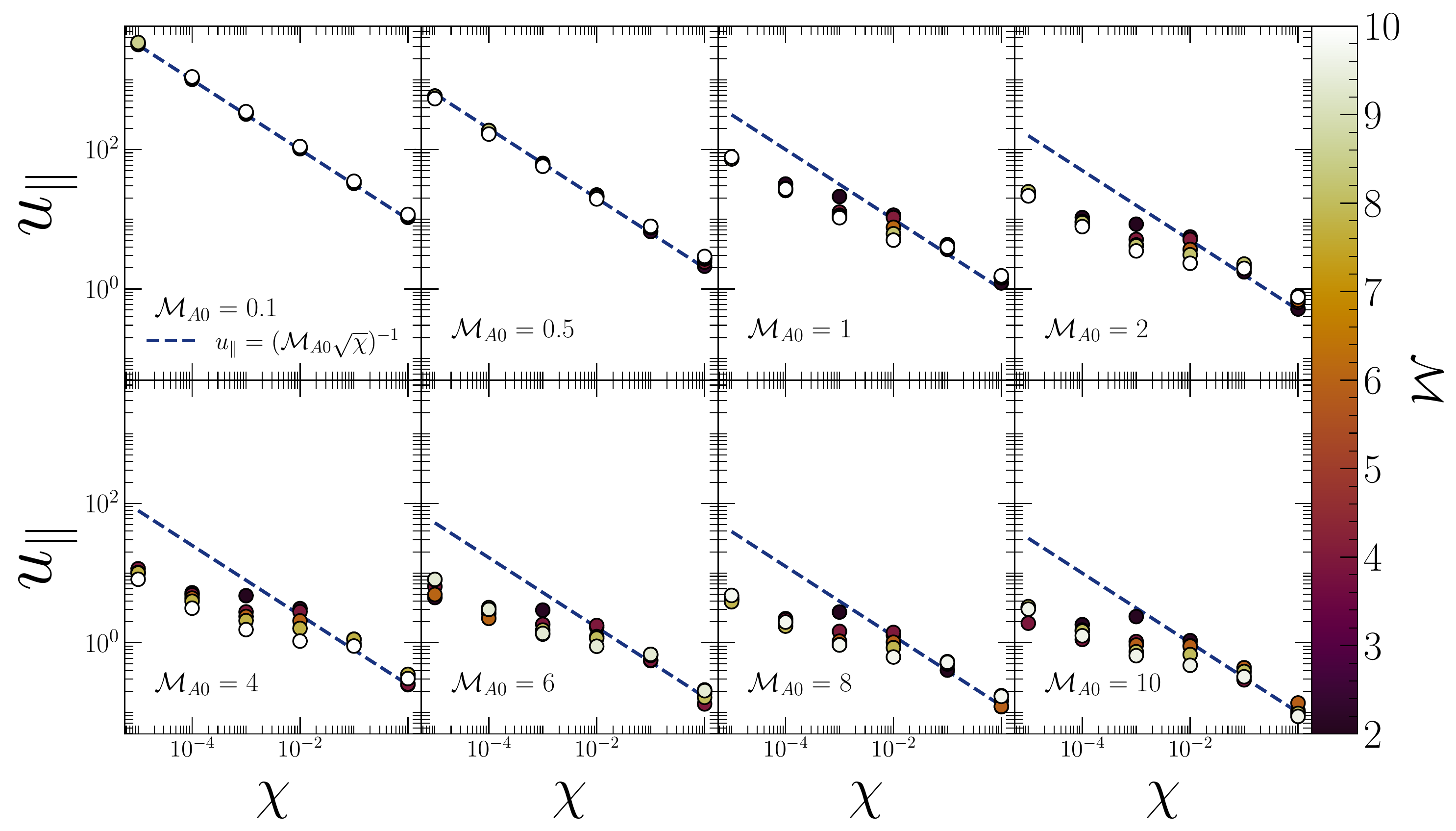}
\caption{ $50^{\rm{th}}$ percentile values from the fits for the parallel drift parameter $u_{\parallel}$ as a function of $\chi$ at fixed $\Mao$. Error bars show the $16^{\rm{th}}$ to $84^{\rm{th}}$ percentile range, but for  most points in the plot this range is so small that the error bars are hidden behind the marker for the $50^{\rm{th}}$ percentile. Panels from left-to-right, top-to-bottom are increasing in $\Mao$. In all panels, points are coloured by $\M$. The dependence on $\M$ is weak since we see no systematic changes in $u_{\parallel}$ with $\M$. The dashed line indicates $u_\parallel = (\Mao \sqrt{\chi})^{-1}$ as expected if $u_\parallel$ is equal to the SCR streaming speed.}
\label{fig:drift}
\end{figure*}

We fit for a drift parameter, $u$, in all simulations, for all spatial dimensions. As expected, we find $u_x \approx u_y \approx 0$ for all trials since we have no preferential SCR direction perpendicular to $\Bo$. Thus we will not discuss these cases further. \autoref{fig:drift} shows the results for $u_\parallel \equiv u_z$ as a function of $\chi$ at each $\Mao$. For comparison, the microphysical streaming speed in our unit system with length measured relative to the turbulence outer scale $\ell_0$ and time measured in units of the turbulent crossing time $\tau$, is
\begin{equation}
    u_{\rm{str}} = \frac{1}{\Mao \sqrt{\chi}}.
\end{equation}
We see that the macroscopic parallel drift rates we measure are close to this value, which is indicated by the dashed line in \autoref{fig:drift}, with the exception of runs with $\Mao \geq 1$ and $\chi < 0.01$. We explore the origin of this deviation in \autoref{Discussion}.

\subsection{Superdiffusivity index: \texorpdfstring{$\alpha$}{Lg}}
\label{sec:alpha}
\begin{figure*}
\centering
\begin{subfigure}[b]{\textwidth}
   \includegraphics[width=0.99\linewidth]{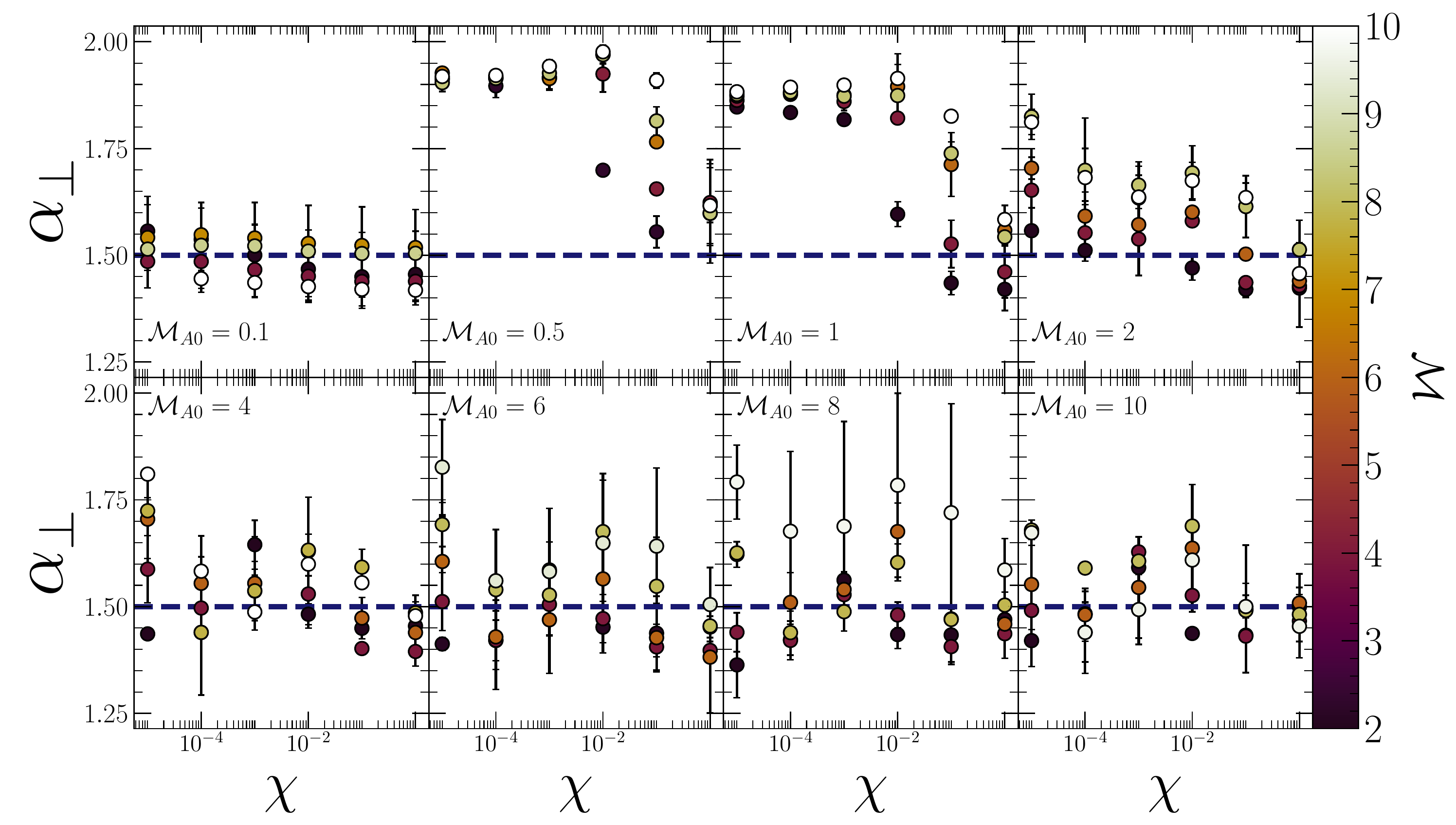}
   \label{fig:mach_perp} 
\end{subfigure}

\begin{subfigure}[b]{\textwidth}
   \includegraphics[width=0.99\linewidth]{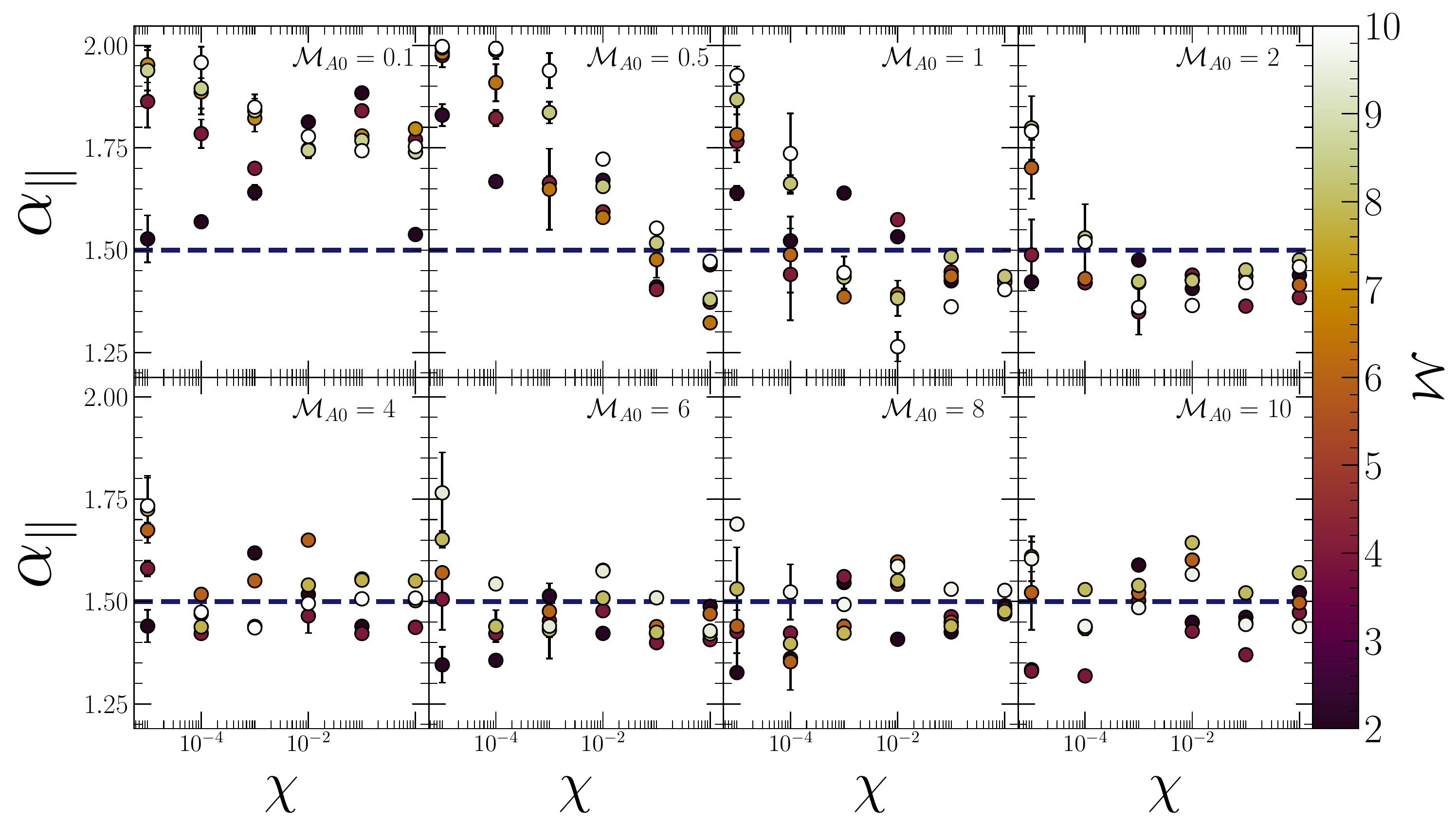}
   \label{fig:mach_par}
\end{subfigure}
\caption{Same as \autoref{fig:drift} but for the perpendicular (top) and parallel (bottom) fractional diffusion indices instead of the drift. In the parallel direction, we plot two points per trial, one corresponding to $\alpha_x$ and the other to $\alpha_y$, the results of our fits in the $x$ and $y$ directions. Note that $\alpha=2$ corresponds to classical diffusion, while $\alpha < 2$ corresponds to superdiffusion. The dashed horizontal lines indicate $\alpha = 1.5$, corresponding to Richardson diffusion (see \autoref{sec:super}).}
\label{fig:a_Alfven}
\end{figure*}

We next examine results for the superdiffusivity index $\alpha$, which represents the fractional power in our generalised diffusion equation (see \autoref{eqn:diffusion}). Physically, $\alpha = 2$ corresponds to classical diffusion and the smaller $\alpha$ becomes compared to 2, the more superdiffusive the system. \autoref{fig:a_Alfven} displays $\alpha$ in the directions parallel ($\alpha_\parallel = \alpha_z$) and perpendicular ($\alpha_\perp = \alpha_x$ or $\alpha_y$ -- we do not differentiate between these two) to the large-scale field as a function of $\chi$ at fixed $\Mao$. All $\alpha$ values lie in the range $ 1.4 \lesssim \alpha < 2$, indicating we have superdiffusion over our entire parameter space. Both parallel and perpendicular fractional diffusion indices exhibit little systematic trend with $\Mao$ or $\chi$, with the exception that in the parallel direction transport seems to approach pure diffusion ($\alpha_\parallel = 2$) for $\Mao \lesssim 1$ and $\chi \lesssim 0.01$. The majority of trials have an $\alpha$ value $\approx 1.5$ for $\Mao \gtrsim 1$. We explore the physical significance of this result in \autoref{Discussion}.

\subsection{Diffusion coefficients: \texorpdfstring{$\kappa$}{Lg}}
\label{sec:kappa}

As with $\alpha$, we separate our results for the diffusion coefficient into values parallel to ($\kappa_\parallel = \kappa_z$) and perpendicular to ($\kappa_\perp = \kappa_x$ or $\kappa_y$) the large scale mean field. These parameters show non-trivial dependence on both $\Mao$ and $\chi$, which we explore below. We reiterate here that our units for $\kappa$ are $\ell_0^{\alpha} / \tau  = \ell_0^{\alpha - 1}c_s \mathcal{M}$, hence any results for $\kappa$ have an intrinsic dependence on the velocity dispersion and driving scale.

\begin{figure*}
\centering
\begin{subfigure}[b]{\textwidth}
   \includegraphics[width=0.99\linewidth]{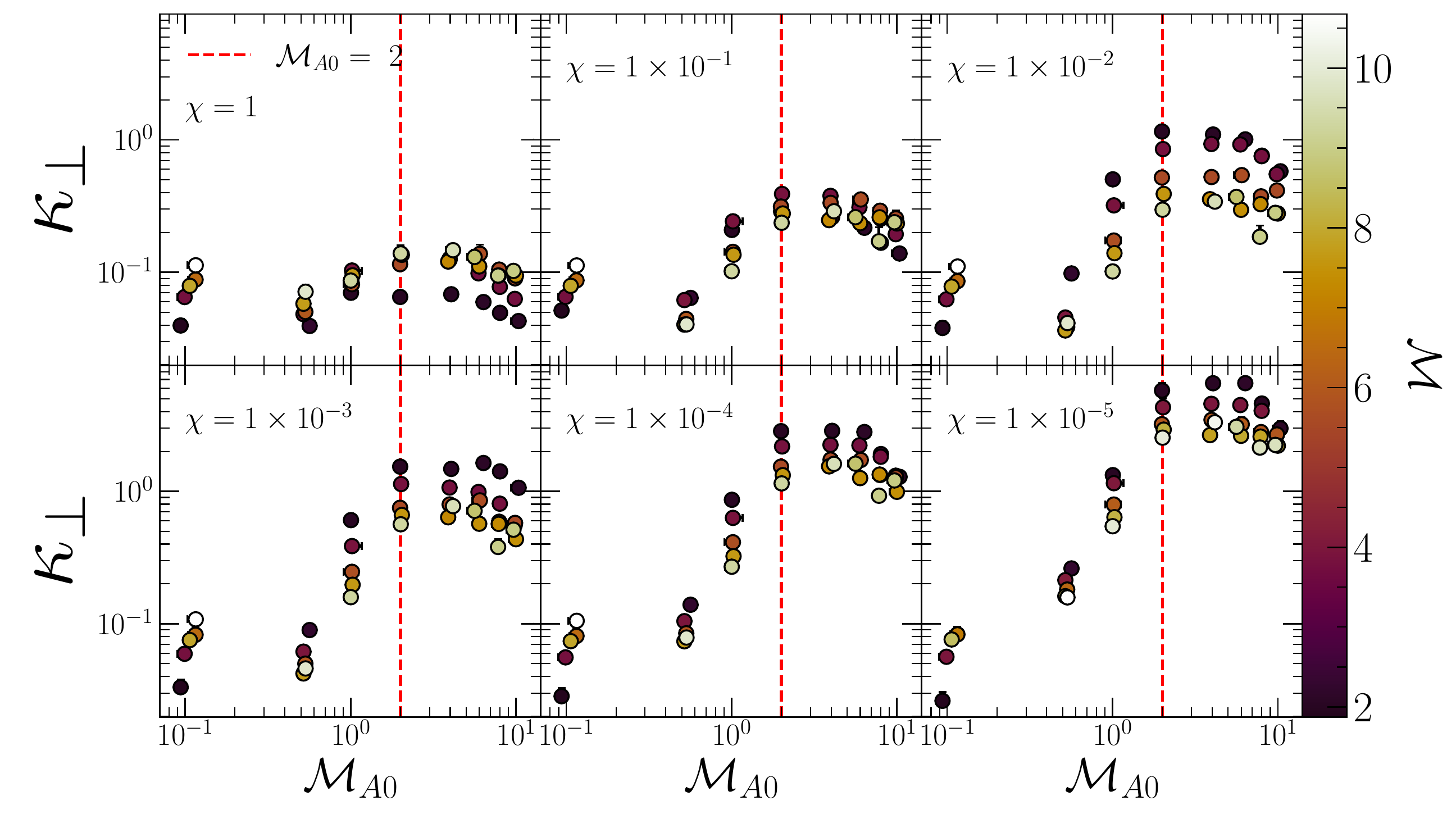}
   \label{fig:mach_perp_a} 
\end{subfigure}
\begin{subfigure}[b]{\textwidth}
   \includegraphics[width=0.99\linewidth]{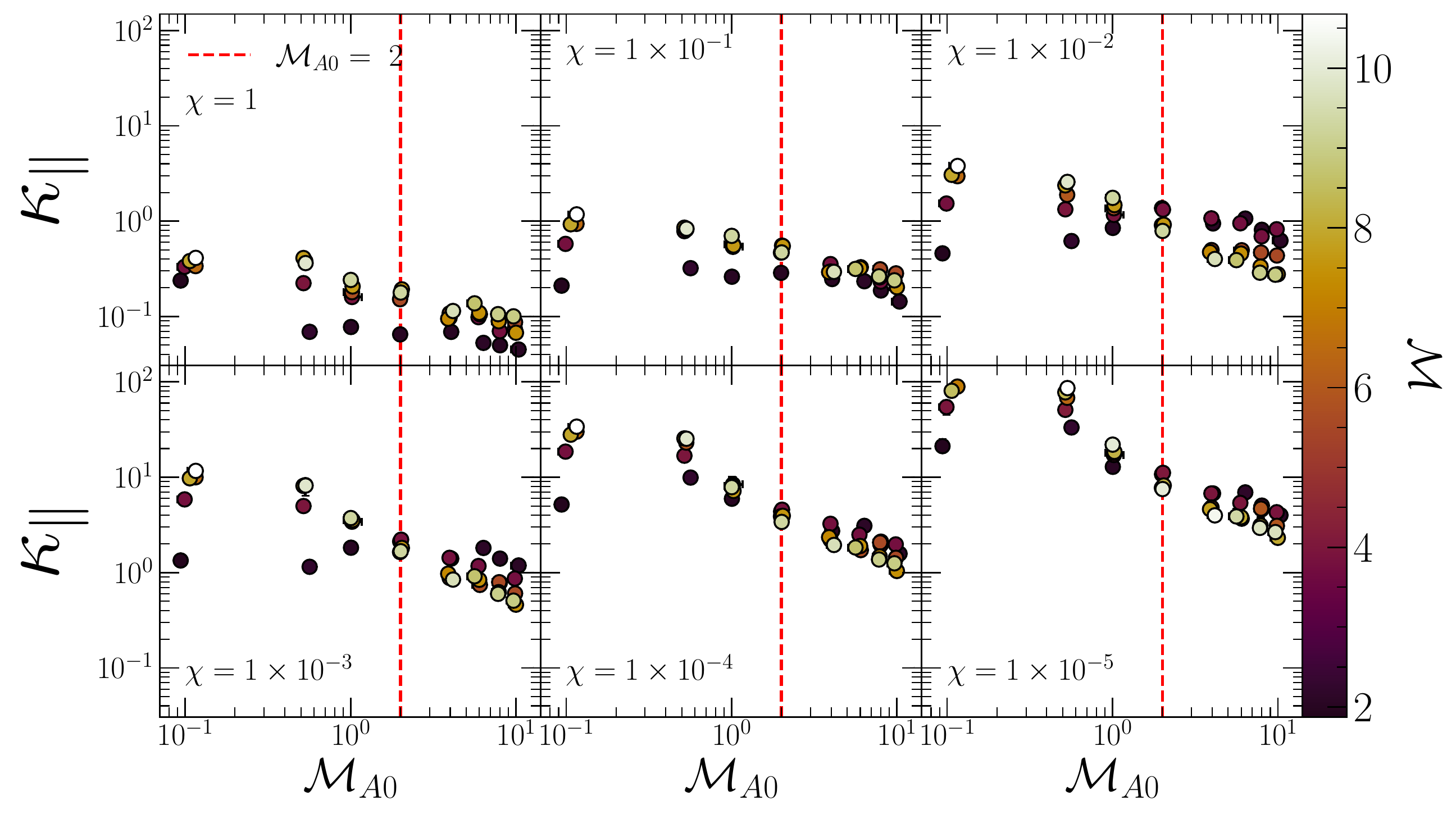}
   \label{fig:mach_par_a}
\end{subfigure}
\caption{Perpendicular (top) and parallel (bottom) diffusion coefficients plotted against Alfv\'en Mach number. The colour bar here indicates the sonic Mach number, while the panels represent decreasing values of $\chi$ (and thus increasing streaming velocities) from left to right, top to bottom. Red vertical dashed lines indicate $\Mao = 2$, marking approximate equipartition between the turbulent and organised parts of the magnetic field.}
\label{fig:Alfven}
\end{figure*}

\autoref{fig:Alfven} shows the fitted diffusion coefficients as a function of Alfv\'en Mach number at fixed $\chi$, with the sonic Mach number shown in colour. Red vertical lines mark $\Mao \approx 2$, the value for which there is approximate equipartition between the turbulent and coherent parts of the magnetic field. We see in the upper plot of $\kappa_{\perp}$ that this value also marks a transition in the behaviour of $\kappa_\perp$: above $\Mao=2$, $\kappa_\perp$ reaches a plateau value such taht further increases in $\Mao$ no longer have any effect on the diffusion rate. The height of this plateau varies inversely with $\chi$. The variation of $\kappa_\parallel$ with $\chi$ is quite different. In the upper panel of \autoref{fig:Alfven}, for $\kappa_\perp$, we see almost a sigmoid like shape, while in the bottom panel for $\kappa_\parallel$ we see a much more linear decrease in $\kappa_{\parallel}$ with $\Mao$, indicating a single power-law relation may be a good fit to these data.

\begin{figure*}
\centering
\begin{subfigure}[b]{\textwidth}
   \includegraphics[width=0.95\linewidth]{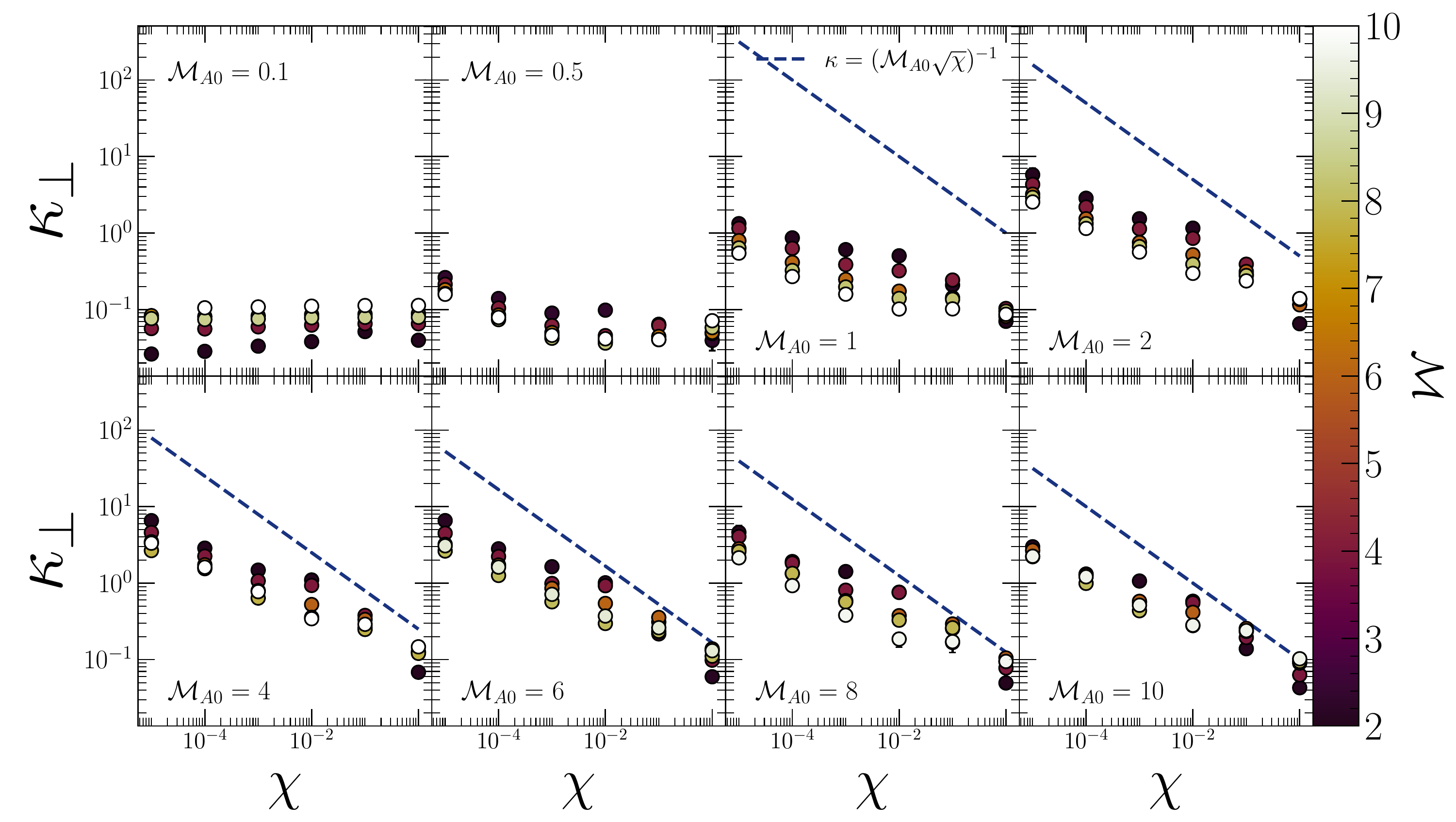}
\end{subfigure}
\begin{subfigure}[b]{\textwidth}
   \includegraphics[width=0.95\linewidth]{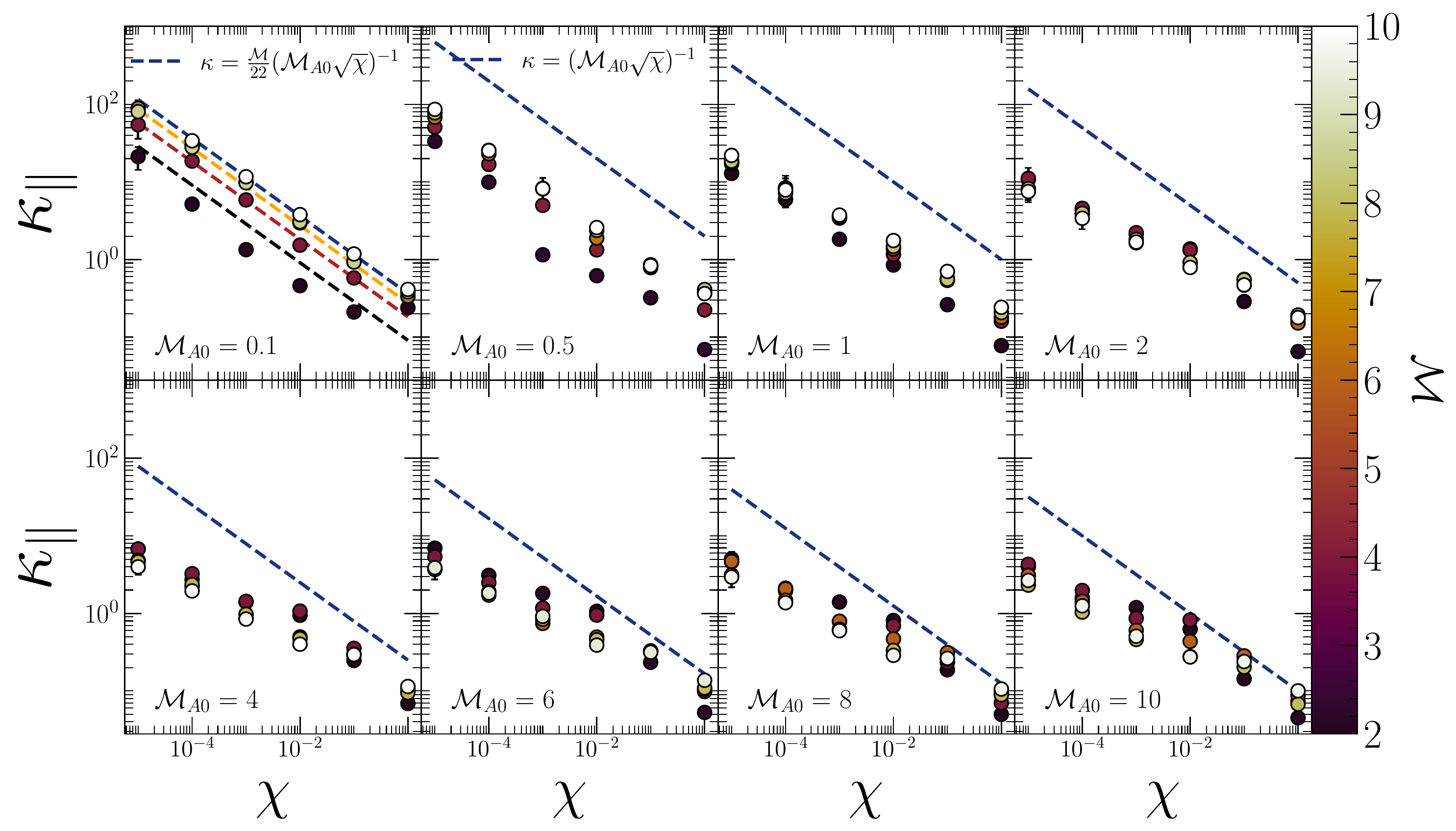}
\end{subfigure}
\caption{Same as \autoref{fig:Alfven}, but now each panel shows fixed $\Mao$, and the plots show the variation of $\kappa$ with $\chi$. Dashed lines show a power-law relation $\kappa = 1 / \Mao \sqrt{\chi}$ in all panels except $\kappa_\parallel$, $\Mao=0.1$, where we instead show $\kappa = (\M / 22) / \Mao \sqrt{\chi}$. The $\kappa\propto 1/\sqrt{\chi}$ scaling is expected if the diffusion coefficient is linearly proportional to the microphysical CR streaming speed, $v_{\rm str} \propto 1/\sqrt{\chi}$.}
\label{fig:chi}
\end{figure*}

In \autoref{fig:chi}, we plot $\kappa_{\parallel}$ and $\kappa_{\perp}$ as a function of $\chi$ at fixed $\Mao$; again, $\M$ is indicated by the colour. For comparison, we also show a power-law relation
\begin{equation}
    \kappa \propto \frac{1}{\sqrt{\chi}},
\end{equation}
motivated by the CR streaming speed being proportional to $1 / \sqrt{\chi}$. At high $\Mao$, we see that $\kappa_\perp$ and $\kappa_\parallel$ behave very similarly, and both approach this powerlaw scaling at high $\chi$. Thus, for a highly tangled fields (large $\Mao$) and relatively slow streaming ($\chi$ close to unity), the diffusion coefficient appears to scale with the streaming speed. However, as the streaming speed increases ($\chi \to 0$), the diffusion rate scales more weakly with $\chi$ that $1/\sqrt{\chi}$, leading to a flatter dependence.

At low $\Mao$ the situation is quite different, and $\kappa_\parallel$ and $\kappa_\perp$ do not scale with $\chi$ in similar ways. We find that $\kappa_{\parallel}$ follows a clear $1 / \sqrt{\chi}$ scaling up to at least $\mathcal{O}(10^2)$ at all $\chi$, while $\kappa_{\perp}$ becomes both small and nearly independent of $\chi$. Instead, $\M$ appears to be the primary factor governing the rate of diffusion.

%%% Commented out, may put in appendix %%%
%\begin{comment}
%\end{comment}
%%%%%%%%%%%%%%%%%%%%%%%%%%%%%%%%%%%%%%%%%%

%%%%%%%%%%%%%%%%%%%%%%%%%%%%%%%%%%%%%%%%%%%%%%%%%%%%%%%%
%%% Discussion section
%%%%%%%%%%%%%%%%%%%%%%%%%%%%%%%%%%%%%%%%%%%%%%%%%%%%%%%%
\section{Discussion}
\label{Discussion}
In this section, we develop a physical picture to help understand the results presented in \autoref{Results}. We begin by presenting an overview of macroscopic diffusion mechanisms, and use this to provide a taxonomy of different diffusive regimes, in \autoref{sec:criptic_diffusion}. We provide fitting formulae to our numerical results, suitable for using in analytic models or simulations that do not resolve ISM turbulence, in \autoref{ssec:fits}.
We discuss superdiffusion and its observational implications in \autoref{sec:super}.
Next, we make a comparison between our results and the literature in \autoref{sec:context}. In \autoref{sec:Limitation}, we discuss the limitations of our study.

%%%%%%%%%%%%%%%%%%%%%%%%%%%%%%%%%%%%%%%%%%%%%%%%%%%%%%%%%%%%%%%%%%%%%
%%% Diffusion in CRIPTIC
%%%%%%%%%%%%%%%%%%%%%%%%%%%%%%%%%%%%%%%%%%%%%%%%%%%%%%%%%%%%%%%%%%%%%

\subsection{Diffusive mechanisms and regimes}
\label{sec:criptic_diffusion}
As described in \autoref{Methods}, our transport equation for CRs (\autoref{eq:FPE}) is purely advective, and includes no explicit diffusion. Nonetheless, we have seen in \autoref{Results} that superdiffusion is a good description of the resulting macroscopic SCR transport. It is therefore of interest to understand what physical mechanisms are responsible for producing the 
{\it effectively}
diffusive behaviour.

\subsubsection{Mechanisms of diffusion}
\label{sssec:mechanisms}

The governing equation of motion for the SCR distribution is \autoref{eq:FPE}, which immediately shows that there are two main channels through which the diffusion may occur: (1) dispersion in $\mathbf{w}$ (i.e., via turbulent advection) and (2) dispersion in $\vstr \mathbf{b}$ (i.e., by either changing $\vstr \propto B/\sqrt{\rho}$ or by changing the direction of the magnetic field $\mathbf{b}$). In principle, dispersion in $\chi$ within a plasma may also be important, but is excluded from this study due to the numerical setup. Based on these two channels for dispersing populations of SCRs, we conclude there are four potential physical mechanisms that will contribute to the macroscopic diffusion:
\begin{enumerate}
    \item magnetic field line tangling and stretching (i.e., fluctuations in $v_{\rm str} \mathbf{b}$ by either changing the magnetic field magnitude, and thus $v_{\rm str}$, or changing the direction of $\mathbf{b}$ with respect to $\Bo$)\footnote{Note that we are using the terminology ``tangling and stretching" instead of the more conventional terminology ``field line random walk" (FLRW). We use this terminology to distinguish between trajectories along static magnetic fields, as considered for example by \citet{yan2008cosmic,snodin2016global}, and time-evolving magnetic fields, such as those in the present study.},
    \item the advection of magnetic field lines (i.e., fluctuations in the component of $\mathbf{w}$ normal to $\mathbf{b}$, which advect the field),
    \item density fluctuations (i.e., fluctuations in $\vstr$ via the density), 
    \item gas flow along field lines (i.e., fluctuations in $\mathbf{w}$ in the direction parallel to $\mathbf{b}$),. 
\end{enumerate}
We explain each of these phenomena in detail below.

%%%%%%%%%%%%%%%%%%%%%%%%%%%%%%%%%%%%%%%%%%%%%%%%%%%%%%%%%%%%%%%%%%%%%
%%% Tangly slinky fields
%%%%%%%%%%%%%%%%%%%%%%%%%%%%%%%%%%%%%%%%%%%%%%%%%%%%%%%%%%%%%%%%%%%%%
\paragraph{Field line tangling and stretching $(\Mao > 2)$.}
\label{sec:tangled}
For $\Mao \gg 1$, where the energy in the ordered part of the magnetic field is much less than the energy in the turbulence, globally the $\mathbf{B}$-field lines become tangled as they are advected by the turbulent motions, and isotropically distributed (e.g., in $k$-space) through the plasma. SCRs will stream along $\mathbf{B}$ and, because the field lines are bent relative to $\Bo$, there will be perpendicular displacement in their position relative to $\Bo$, leading to diffusion in the perpendicular direction. Even though the fields are statistically isotropic on large-scales, field line tangling also results in inhomogeneous magnetic field amplitudes in space and time \citep[see Figure~1 in ][]{beattie2020magnetic}. Hence, SCRs separated on larger scales than the magnetic correlation length will experience different magnetic field amplitudes, giving rise to variations of $\vstr \propto B/\sqrt{\rho}$ between the populations, producing parallel diffusion as well. The final result is effective macroscopic diffusion, globally. This mechanism can also affect the net streaming speed, so that, for $\chi \ll 1$, $u_{\parallel} \ll \vstr$. We can understand this as a result of field line tangling as well: over a fixed time $\Delta t$, SCRs moving along bent field lines will be displaced less along the large scale $\Bo$ field direction, $\Delta \ell$, than they would had the field lines been straight. This effect is large when $\chi \ll 1$, because in this case $\vstr \gg w$, and the field lines do not have time to move significantly as CRs stream down them, so CRs travel along curved paths, lowering their mean rate of progress down the field, $u_{\parallel}$. By contrast, when $\chi \sim 1$, $\vstr \la w$, and the field lines have significant time to move as CRs stream down them. Thus the net streaming speed becomes sensitive not to the instantaneous shape of the field lines, but to their time-averaged shape, which remains aligned with $\Bo$. This explains why field line tangling produces $u_\parallel \ll \vstr$ for $\chi \ll 1$, and $u_\parallel \sim \vstr$ for $\chi \sim 1$.

%%%%%%%%%%%%%%%%%%%%%%%%%%%%%%%%%%%%%%%%%%%%%%%%%%%%%%%%%%%%%%%%%%%%%
%%% FLRW
%%%%%%%%%%%%%%%%%%%%%%%%%%%%%%%%%%%%%%%%%%%%%%%%%%%%%%%%%%%%%%%%%%%%%
\paragraph{The advection of magnetic field lines $(\Mao < 2)$.}
\label{sec:random_walk}
%% Add limiting case to give better context to whats going on
% The curvature of magnetic field lines can be determined by $\Mao$ \citep{yuen2020curvature}. In general, as $\Mao \to 0$, the field lines become perfectly aligned with $\Bo$, with only very small fluctuations, (e.g., $\Exp{\delta B^2}/B_0^2 = \mathcal{M}_{\rm A0}^{4}/4$, for $\Mao \leq 2$; \citealt{beattie2020magnetic,Beattie2022_inprep_energybalance}). 
%When $\mathbf{B}$-field lines are advected by the velocity fluctuations, SCRs that are tied to those field lines move with them. This process is referred to as field line random walk (FLRW) \citep{jokipii1968random}. 
In $\Mao < 2$ plasmas, the ordered $\mathbf{B}$-fields have %much 
more energy than is contained in the turbulence.\footnote{Note that for $\Mao = 2$ the fluctuating and large-scale field are in energy equipartition, $\Exp{\delta B^2}/B_0^2 = 1$ \citep{beattie2020magnetic}.} Because of this, field lines resist bending via the magnetic tension force, and the turbulence can only shuffle the field lines in  motions perpendicular to $\Bo$. In this regime, due to flux freezing and the extremely rigid $\mathbf{B}$-field, the velocity is unable to produce strong diagonal modes and can only maintain either flows along the field lines, $u_{\parallel}$, or perpendicular to the field lines, $u_{\perp}$, creating self-organised, solid-body ($u_{r}\propto r$, where $r$ is the radial coordinate with the vortex core at the origin) vortices in the plane perpendicular to $\Bo$ \citep[see Figure~10 and \S7.1 in ][]{Beattie2021_spdf}. The dynamical timescale of the outer vortex rotation is set by the correlation time of the turbulent driving $\tau$ (\autoref{eq:turnover_time}) which scales inversely with $\M$. In such a system, there will be perpendicular dispersion in the SCRs' positions based on where they are in the vortex plane.
%When there is significant amounts of curvature in the $\mathbf{B}$-field lines, i.e., $\Mao > 2$, the advection will provide both parallel and perpendicular dispersion. In contrast, when the field lines are straight, $\Mao \lesssim 2$, we expect this process to produce diffusion in the perpendicular direction only, since only velocities perpendicular to the field move the field lines. To understand the distinction between FLRW and field line tangling, consider what happens in the limit $v_{\rm str}\to 0$. In this case field line tangling would have no effect, because CRs never move relative to the field lines; however, CRs could still be transported by FLRW.

%%%%%%%%%%%%%%%%%%%%%%%%%%%%%%%%%%%%%%%%%%%%%%%%%%%%%%%%%%%%%%%%%%%%%
%%% Gas flow along field lines
%%%%%%%%%%%%%%%%%%%%%%%%%%%%%%%%%%%%%%%%%%%%%%%%%%%%%%%%%%%%%%%%%%%%%
\paragraph{Gas flow along the field.}
\label{sec:vel_stream}
As the $\mathbf{B}$ field is advected with the gas so will the SCRs travelling along them, and hence the total SCR velocity is a vector sum of the streaming velocity plus the gas velocity (as indicated in \autoref{eq:FPE}). 
%Velocity fluctuations perpendicular to the local magnetic field result in either  for sub-Alfv\'enic turbulence (as discussed in \autoref{sec:FLRW}), or field line tangling for super-Alfv\'enic turbulence (discussed in \autoref{sec:tangled}). However,
As SCRs are transported along strong field lines, velocities parallel to the field lines will act to either slow them down (when the velocity and the streaming direction are antiparallel) or speed them up (when the velocity and streaming direction are parallel). This process is represented by the $\mathbf{v}$ term in \autoref{eq:FPE}. We see from \autoref{eq:FPE} that we only expect significant contributions to diffusion via this process if $|\mathbf{v}|$ is comparable to $|\mathbf{w}|$ (e.g., for large $\chi$). Since the velocity field varies in space and time, the process causes dispersion in the position of SCRs propagating at different places at different times, and therefore acts like a diffusive process. For sub-Alfv\'enic turbulence, where there is reasonable alignment between $\mathbf{B}$ and $\Bo$, this mechanism produces almost purely parallel diffusion. For super-Alfv\'enic turbulence, where $\mathbf{B}$ is not %necessarily 
preferentially
aligned with $\Bo$, the diffusion contribution from the velocity fluctuations is likely isotropic.

%%%%%%%%%%%%%%%%%%%%%%%%%%%%%%%%%%%%%%%%%%%%%%%%%%%%%%%%%%%%%%%%%%%%%
%%% Density fluctuations
%%%%%%%%%%%%%%%%%%%%%%%%%%%%%%%%%%%%%%%%%%%%%%%%%%%%%%%%%%%%%%%%%%%%%
\paragraph{Density fluctuations.}
\label{sec:dens_fluc}
Gas density fluctuations in compressible turbulence are extremely inhomogenous, and are not isotropic when the large-scale field is strong \citep{Beattie_2020_filaments_and_striations}\footnote{Note that even though the density fluctuations are not isotropic, i.e., do not exhibit perfect rotational symmetry in $k$-space, they do however qualitatively support a rotational symmetry around $\Bo$, hence the $k$-modes form a set of nested ellipsoids. The alignment of the semi-minor and semi-major axes depends upon the value of $\M$ but are always along $\Bo$ \citep{Beattie_2020_filaments_and_striations}.}. Since $\vstr \propto B/\sqrt{\rho}$, inhomogeneities in the gas density can produce macroscopic diffusion in the positions of SCRs, regardless of the value of $\Mao$, as temporally and/or spatially separated SCRs populations experience different density fluctuations along their trajectories \citep{Beattie2022_va_fluctuations}. For $\Mao < 1$ turbulence, when the magnetic field fluctuations are negligible, \citet{Beattie2022_va_fluctuations} showed that, in the absence of strong advection, density fluctuations formed from gas flows along the magnetic fields solely determine the macroscopic diffusion in $\ell_{\parallel}$. For $\Mao > 1$, both the $B$-field fluctuations and correlations between the $B$ and $\rho$ grow and contribute to setting the fluctuations in $\vstr$. 

%Since $v_{\rm{str}}$ is dependent on $\chi$ we note that local variations in $\chi$ along field lines would act in an identical way to mass density fluctuations. In this study we have assumed a constant $\chi$ throughout the plasma for each trial. However, this assumption may not be realistic, a topic which we return to in  \autoref{Conclusions}.

%%%%%%%%%%%%%%%%%%%%%%%%%%%%%%%%%%%%%%%%%%%%%%%%%%%%%%%%%%%%%%%%%%%%%
%%% Anisotropic
%%%%%%%%%%%%%%%%%%%%%%%%%%%%%%%%%%%%%%%%%%%%%%%%%%%%%%%%%%%%%%%%%%%%%
\subsubsection{Regimes of turbulent diffusion}
\label{sec:regimes}
\begin{figure*}
    \centering
    \includegraphics[width=\textwidth]{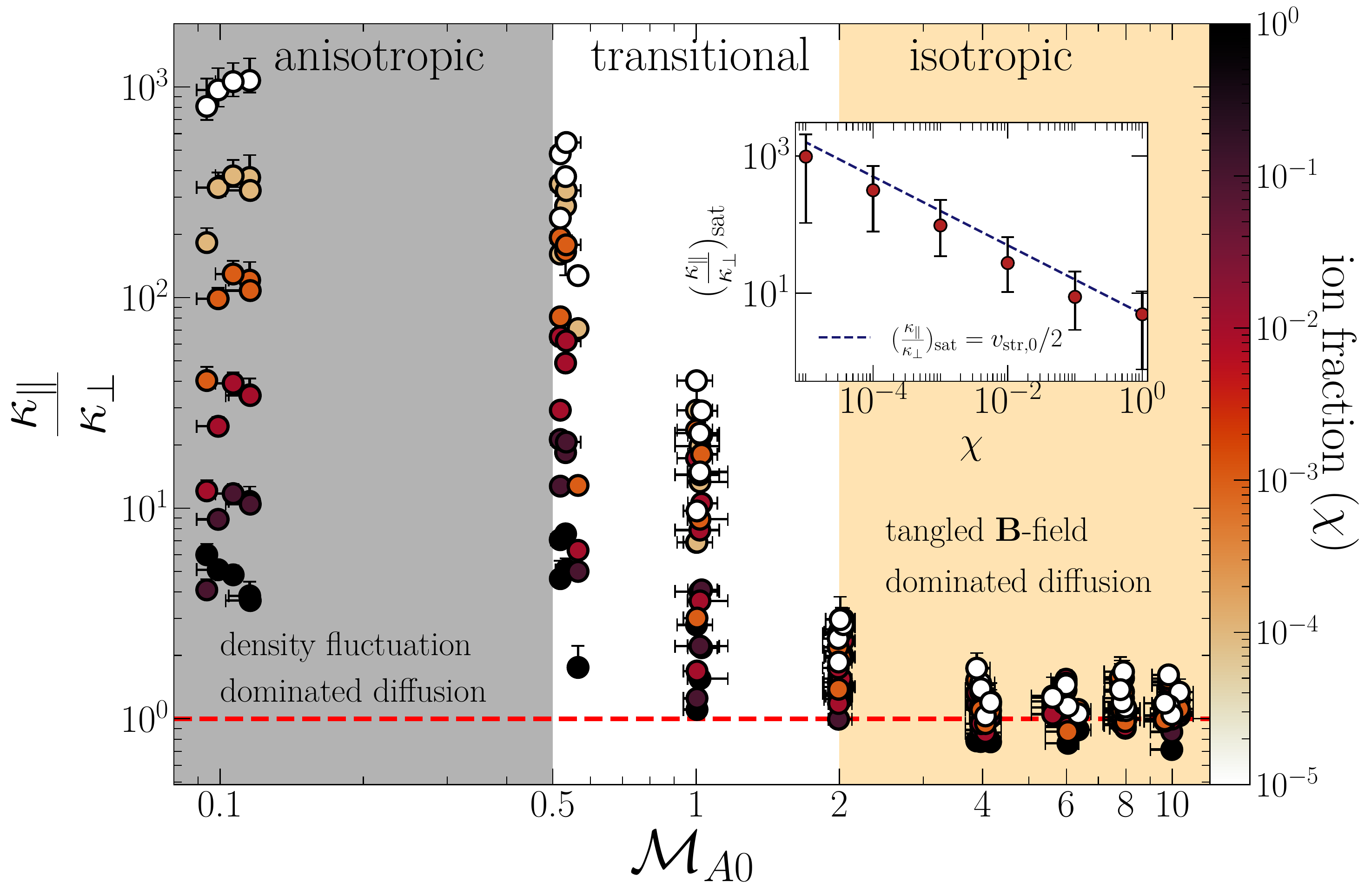}
   \caption{Ratio of parallel and perpendicular diffusion coefficients as a function of $\Mao$ for all trials; the ion fraction $\chi$ is shown in the colour bar. Note that, particularly for the runs with a target $\Mao = 01$, the actual $\Mao$ values scatter slightly around the target because the actual velocity dispersion produced by our driven turbulence simulations fluctuates slightly relative to the target value we select by turning the driving rate. We highlight three distinct regions of $\Mao$, characterised by the dominance of different diffusion mechanisms, which we term anisotropic ($\Mao\lesssim 0.5$), transitional ($0.5 \lesssim \Mao \lesssim 2$) and isotropic ($\Mao\gtrsim 2$). In the anisotropic region we have a clear difference between the rates of perpendicular and parallel diffusion. The level of anisotropy in this region is governed by $\chi$, which we illustrate in the inset plot showing $\kappa_{\parallel} / \kappa_{\perp}$ plotted against $\chi$ for the simulations with $\Mao=0.1$, together with the simple scaling $\kappa_\parallel/\kappa_\perp = v_{\rm str}/2$; the data points shown are averages over the runs with different $\mathcal{M}$, with the error bars showing the $1\sigma$ scatter about this average.}
    \label{fig:all_in_one}
\end{figure*}

We have seen empirically from the results presented so far, and theoretically from the discussion in \autoref{sssec:mechanisms}, that macroscopic SCR diffusion works very differently in the $\Mao \lesssim 2$ and $\Mao > 2$ regimes (noting we are operating in the supersonic regime $\mathcal{M} \geq 2$). We can see this clearly if we plot the ratio of the parallel and perpendicular diffusion coefficients, which we do in \autoref{fig:all_in_one}. We indicate three distinct regimes of diffusion (strongly related to the sub- and super- Alfv\'enic regimes). The anisotropic regime is located at $0 \lesssim \Mao \lesssim 0.5$, transitional, at $0.5 \lesssim \Mao \lesssim 2$, and then finally the isotropic, at $\Mao \gtrsim 2$.
A clear %result from this 
implication here
is the importance of $\Mao$ for SCR transport. Our results suggest both the dominant physical mechanisms and diffusion coefficients will be strongly dependent on whether the SCRs are being transported in a sub-Afv\'enic or super- Alfv\'enic plasma.

\paragraph{Anisotropic regime.}
\label{sec:FLRW}
The anisotropic diffusion regime, which occurs when $\Mao$ is small, is characterised by $\kappa_\parallel / \kappa_\perp \gg 1$, which can be seen in \autoref{fig:chi} and \autoref{fig:all_in_one}. In this regime, $\kappa_\perp$ is nearly independent of $\chi$, but scales with $\M$, increasing from $\approx 0.02$ to $\approx 0.1$ as $\M$ increases from 2 to 8. By contrast $\kappa_{\parallel}$ varies with both $\M$ and $\chi$; an approximate empirical fit to the data is $\kappa_{\parallel} = \M /(22 \  \Mao\sqrt{\chi})$, which shows $\kappa_\parallel$ scales close to linearly with $\M$. We can interpret these results in light of the physical mechanisms discussed in \autoref{sec:criptic_diffusion}. In the sub-Alfv\'enic regime, the $\mathbf{B}$-field lines are extremely resistant to bending, so field line tangling (\autoref{sec:tangled}) is negligible. Diffusion in the perpendicular direction will therefore arise solely from field line advection (\autoref{sec:random_walk}), while parallel diffusion is produced only by gas flow along the field (\autoref{sec:vel_stream}) and density fluctuations (\autoref{sec:dens_fluc}). 

First consider the perpendicular direction. Since field line advection is the only mechanism, it is immediately clear why $\chi$ does not matter: transport depends only on the motion of the background gas (which is not sensitive to $\chi$), not on the flow of CRs relative to it (which is). To first order we might expect $\kappa_\perp$ to be independent of $\mathcal{M}$ as well, since we are working in a unit system where times are already normalised to the turbulent eddy turnover time. The fact that we find a weak increase of $\kappa_\perp$ with $\M$ is probably a sign of the vortices that advect the field lines being more disordered, and thus more effective at diffusing the field lines and the CRs attached to them, as $\M$ increases. Nonetheless, this effect is relatively small.

Now consider the parallel direction, where diffusion comes from random modulation of the streaming speed by fluctuations in the gas density (\autoref{sec:dens_fluc}) and velocity (\autoref{sec:vel_stream}). We therefore expect  the diffusion coefficient to increase with the strength of the modulations, and thus with $\M$, and with the underlying streaming speed that is being modulated (and thus with $\sqrt{\chi}$). This is precisely what we observe. As with the perpendicular direction, the scaling with the streaming speed, and thus with $\chi$, is dominant, so that the ratio of $\kappa_\parallel$ and $\kappa_\perp$ depends only extremely weakly on $\mathcal{M}$. Indeed, as the inset plot in \autoref{fig:all_in_one} shows, the $\kappa_{\parallel} / \kappa_{\perp}$ ratio in the low-$\Mao$ limit is consistent with the simple scaling
\begin{equation}
\lim_{\Mao \rightarrow 0}\frac{\kappa_{\parallel}}{\kappa_{\perp}} = \frac{v_{\rm str,0}}{2} = \frac{1}{2\Mao \sqrt{\chi}},\label{eqn:inset}
\end{equation}
where $v_{\rm str,0}$ is the (dimensionless) streaming speed at the average simulation density and magnetic field. As $\kappa_{\perp}$ is independent  of $\chi$ in this regime, with a systematic scatter set by $\M$ (see top left panel of \autoref{fig:chi}) this shows that $\kappa_{\parallel} \propto v_{\rm str,0} / 2$. In this regime \citet{Beattie2022_va_fluctuations} finds that $\kappa_{\parallel} \propto (1/4)\sigma_s^2 v_{\rm str,0}\ell_{\rm{cor},\rho/\rho_0}$, where $\sigma_s^2$ is the logarithmic gas density variance and $\ell_{\rm{cor},\rho/\rho_0}$ is the gas density correlation scale. By substituting this model into \autoref{eqn:inset}, $\kappa_{\perp} \propto (1/2)\sigma_s^2 \ell_{\rm{cor},\rho/\rho_0}$, i.e., in our dimensionless units of large-scale turnover times and velocities, $\kappa_{\perp}$ is set by the density fluctuations. Because $\sigma_s^2 \sim \ln(1+\M^2)$ \citep{Federrath2008,federrath2010comapring,Molina2012_dens_var,beattie2021multishock}, this explains the scatter in $\M$ that we see in the top left panel of \autoref{fig:chi}. We postpone a more detailed comparison between \citet{Beattie2022_va_fluctuations} theory and our results for a future study. 

%%%%%%%%%%%%%%%%%%%%%%%%%%%%%%%%%%%%%%%%%%%%%%%%%%%%%%%%%%%%%%%%%%%%%
%%% Isotropic
%%%%%%%%%%%%%%%%%%%%%%%%%%%%%%%%%%%%%%%%%%%%%%%%%%%%%%%%%%%%%%%%%%%%%
\paragraph{Transition and isotropic regimes.}
\label{sec:Isotropic}
The defining feature of the isotropic regime is  $\kappa_\parallel / \kappa_\perp$ approaching unity, which can be seen in \autoref{fig:all_in_one}. The equality between the parallel and perpendicular diffusion coefficients is driven by two opposite trends in the transition regime: sharp increases in $\kappa_{\perp}$ and sharp increases in $\kappa_\parallel$ as $\Mao$ increases from $\approx 0.5$ to $\approx 2$ (\autoref{fig:Alfven}), at which point the coefficients converge to $\approx (\Mao\sqrt{\chi})^{-1}$ at $\Mao\approx 2$. At $\Mao \gtrsim 2$, the dependence on $\Mao$ disappears and $\kappa$ depends only on $\chi$. For $\chi$ close to unity, this dependence is approximately $\kappa_\perp \approx \kappa_\parallel \propto \chi^{-1/2}$, but the dependence flattens at $\chi \lesssim 0.01$, where the diffusion coefficients reach a maximum of $\approx 1-2$.

As in the anisotropic regime, we can interpret these numerical results in terms of the physical mechanisms introduced in \autoref{sec:criptic_diffusion}. When $\Mao$ is large, the field lines are easily bent by the flow, and we expect the dominant process in this regime to be field line tangling (\autoref{sec:tangled}). In this regime, an increase in the speed of travel along the field lines (i.e., a decrease in $\chi$) results in a corresponding increase in the rate of diffusion in all directions, with $\kappa_\perp \approx \kappa_\parallel \propto v_{\rm str} \propto \chi^{-1/2}$, as we observe. This is similar to the scaling of $\kappa_\parallel$ in the anisotropic regime: when diffusion acts by random modulation of the streaming speed, the result is a diffusion coefficient that scales linearly with the streaming speed.

However, this does not explain the flattening in $\kappa_\perp$ and $\kappa_\parallel$ at low $\chi$, which prevents them from exceeding $\approx 1-2$, corresponding to diffusing a distance of order $\ell_0$ per time $\tau$. We hypothesise that this saturation is due to the rate at which field lines become space filling inside our domain. That is, no matter how fast SCRs may be travelling along $\mathbf{B}$-field lines, the rate of diffusion is ultimately limited by the timescale on which the tangled field lines are able to explore all space in the box.  %Our results imply that the space explored by the SCRs streaming along bent $\mathbf{B}$-field lines will become space-filling after one turnover time when $\Mao \geq 2$. A caveat to this is the slight upturn in $\kappa_{\parallel}$ and $\kappa_{\perp}$ seen in \autoref{fig:chi} for $\chi = 10^{-5}$. The upturn is more pronounced in certain $\Mao$ trials. 

\subsection{Fitting formulae}
\label{ssec:fits}
As discussed in \autoref{sec:intro}, one of the primary motivations for our work is to provide an effective theory for CR transport that can be used in cosmological or galactic-scale simulations that do not resolve turbulence in the ISM. To facilitate this, in this section we construct a series of models to calculate CR diffusion coefficients given values for $\Mao$ and $\chi$; we omit $\M$ since its effects are small within our system units of time = $\tau$. The intended use for these models is much the same as in large eddy simulations: one can measure the plasma parameters at the minimum resolved scales, and use these in the formulae provided below to assign an effective subgrid diffusion coefficient for CRs due to the unresolved turbulent structure and flow; we discuss below how to treat superdiffusion approximately in such a framework.
%Note that for all these models the units are $\ell_0^{\alpha} / \tau$, where $\ell_0$ and $\tau$ come from the simulation in which the sub-grid model is to be added. 
Since we have seen that there are two general regimes for CR transport, corresponding to $\Mao\ll 1$ and $\gg 1$, and that the parameters describing transport are relatively flat in each of these two regimes, we fit all quantities using a generic functional form
\begin{equation}
    f(\Mao,\chi) = p_0 \chi^{p_1} + p_2 \chi^{p_3} \left\{
    \frac{\tanh\left[p_4 \left(\log\Mao - p_5\right)\right] + 1}{2}
    \right\}.
    \label{eq:fit_func}
\end{equation}
The function in curly braces has the property that it goes to zero for when $\Mao \to 0$ (for positive $p_4$) and to unity for $\Mao\to \infty$, which provides the two flat plateaus at low and high $\Mao$ that we have observed. The parameters $p_4$ and $p_5$ control the steepness and location of the transition between the two plateaus, respectively; $p_0$ and $p_1$ provide the normalisation and dependence on $\chi$ for one plateau, while $p_2$ and $p_3$ serve the same purpose for the other plateau.

\begin{table}
    \centering
    \begin{tabular}{l| r@{}l r@{}l r@{}l}
    \hline\hline
        & \multicolumn{6}{c}{Fit quantity} \\
        Parameter & \multicolumn{2}{c}{$\kappa_\parallel/\kappa_\perp$} & \multicolumn{2}{c}{$\kappa_\perp$} & \multicolumn{2}{c}{$u_\parallel/v_{\rm str}$} \\ \hline
        $p_0$ & $1.077$\,    & $\pm\, 0.069$ & $0.0541 $\,    & $\pm\, 0.0050$ & $1.546 $\,    & $\pm\, 0.060$ \\
        $p_1$ & $-0.0175$\,    & $\pm\, 0.0097$ & $-0.017$\,    & $\pm\, 0.016$ & $0.223$\,    & $\pm\, 0.0058$\\
        $p_2$ & $5.65$\,    & $ \pm\, 0.69$ & $0.0804$\,    & $\pm\, 0.0074$ & $0.306$\,    & $\pm\, 0.071$\\
        $p_3$ & $-0.403$\,    & $ \pm\, 0.015$ & $-0.324$\,    & $\pm\, 0.011$ & $-0.110$\,    & $\pm\, 0.024$ \\
        $p_4$ & $-5.94$\,    & $\pm\, 0.42$ & $5.59$\,    & $\pm\, 0.62$ & $-7.1$\,    & $\pm\, 1.9$ \\
        $p_5$ & $-0.201$\,    & $\pm\, 0.022$ & $0.074$\,    & $\pm\, 0.019$ & $-0.132$\,    & $\pm\, 0.041$
         \\ \hline \hline
    \end{tabular}
    \caption{Best-fit parameters $p_0$ - $p_5$, with uncertainties, for fits of the quantities indicated to the functional form given by \autoref{eq:fit_func}.}
    \label{tab:fit_param}
\end{table}

\begin{figure}
    \centering
    \includegraphics[width=0.47\textwidth]{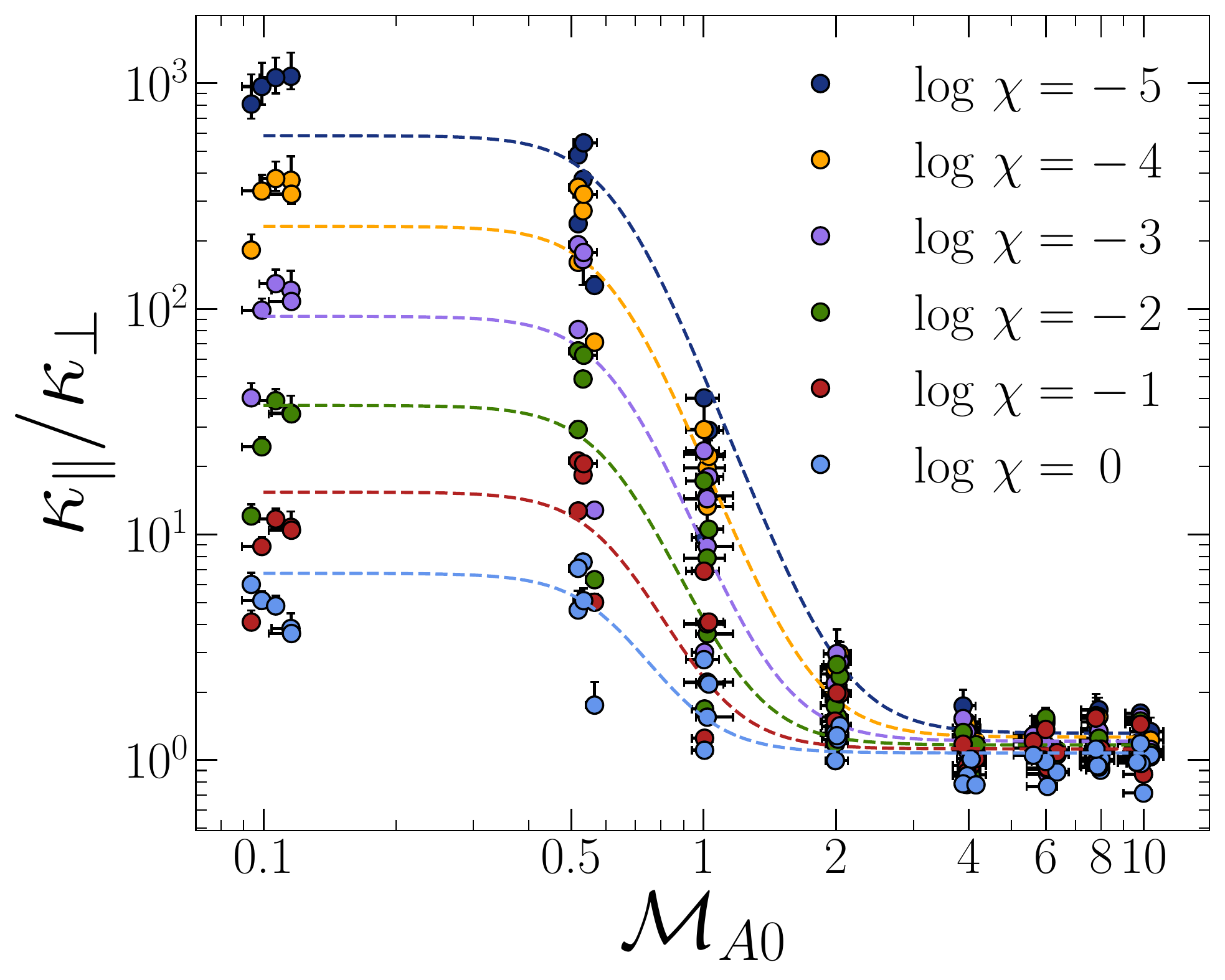}
    \caption{Numerical results for $\kappa_{\parallel} / \kappa_{\perp}$ as a function of $\Mao$ (circles, with colour indicating $\chi$) compared to our empirical fitting formula (\autoref{eq:fit_func}, using the parameters from \autoref{tab:fit_param}). The different dashed lines show the fitting formula evaluated for $\log\chi = -5$ (top) to $0$ (bottom) in steps of $\Delta\log\chi = 1$.}
    \label{fig:iso}
\end{figure}

\begin{figure}
    \centering
    \includegraphics[width=0.47\textwidth]{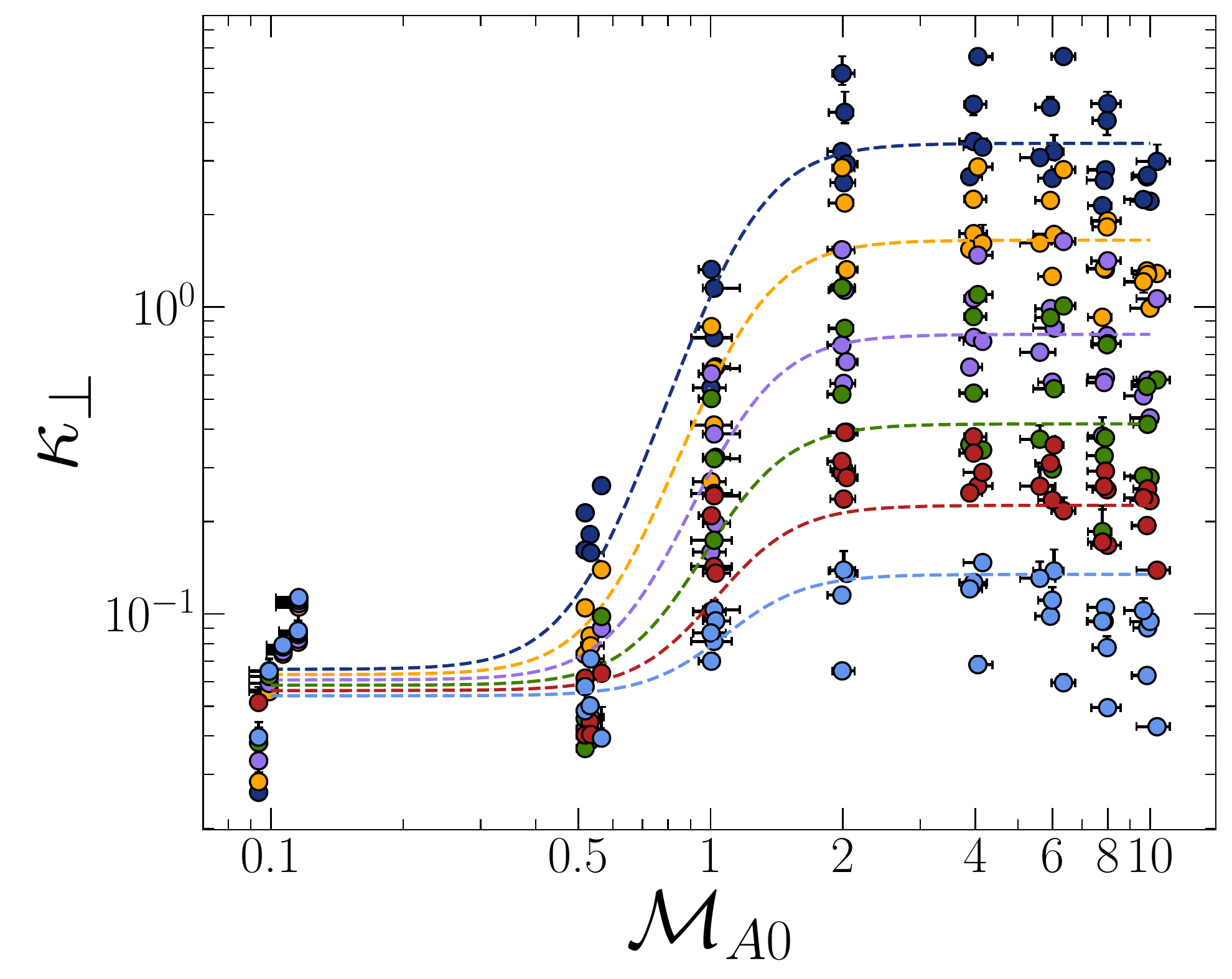}
    \caption{Similar to \autoref{fig:iso} but for $\kappa_{\perp}$.}
    \label{fig:perp_fit}
\end{figure}

\begin{figure}
    \centering
    \includegraphics[width=0.47\textwidth]{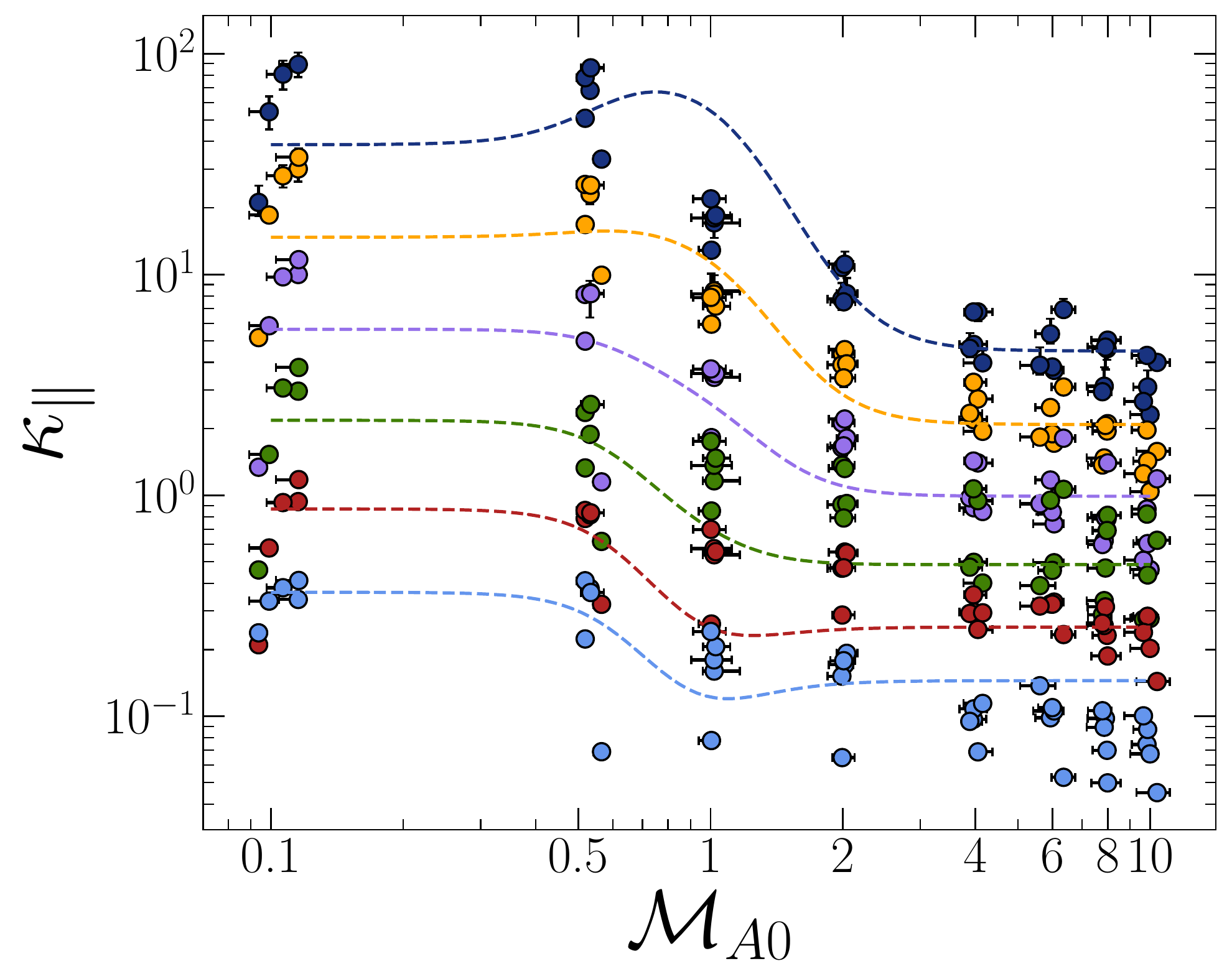}
    \caption{Similar to \autoref{fig:iso} but for $\kappa_{\parallel}$. Dashed lines show models, computed by multiplying our fits for $(\kappa_\parallel/\kappa_\perp)$ and $\kappa_\perp$.}
    \label{fig:par_fit}
\end{figure}

\begin{figure}
    \centering
    \includegraphics[width=0.47\textwidth]{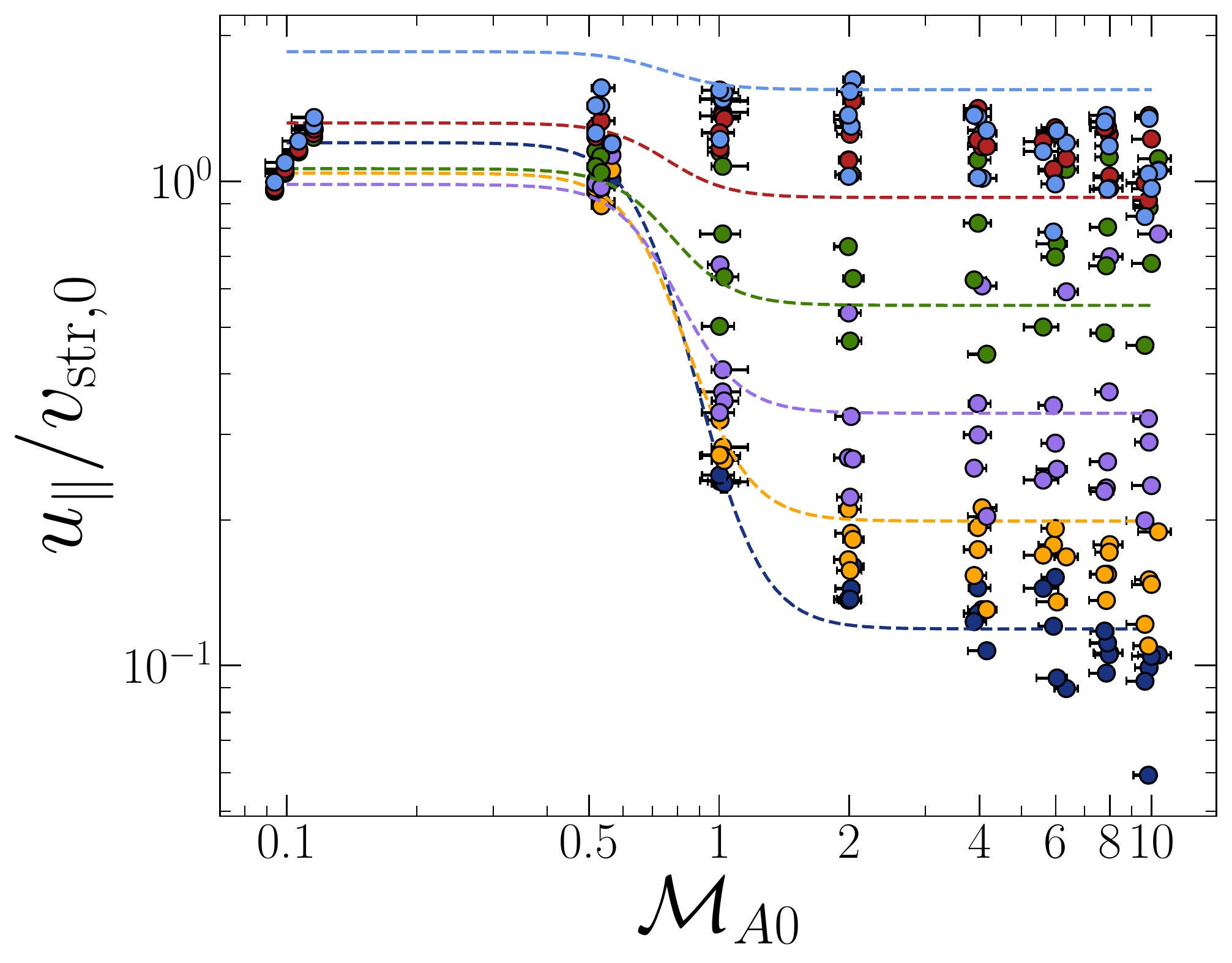}
    \caption{Similar to \autoref{fig:iso} but for $u_{\parallel}$ normalised by the mean small-scale streaming speed,  $v_{\rm{str},0}= (\Mao\sqrt{\chi})^{-1}$.}
    \label{fig:drift_fit}
\end{figure}

We begin by providing a fit for $\kappa_\parallel/\kappa_\perp$, which quantifies the anisotropy of the diffusion. We perform a simple non-linear least squares fit of our data for $\log(\kappa_\parallel/\kappa_\perp)$ from all our simulations, weighting them all equally, to a functional of the form $\log f(\Mao,\chi)$, where $f$ given by \autoref{eq:fit_func}. We report the best-fit parameters and their uncertainties in \autoref{tab:fit_param}, and we plot our fit against the data in \autoref{fig:iso}, which shows that the fit captures the basic trends well. We repeat this process for $\kappa_\perp$ and for $u_\parallel/v_{\rm str,0}$, where $v_{\rm str,0} = 1/\Mao\sqrt{\chi}$ is the mean small-scale streaming speed. We report our fit parameters for these quantities in \autoref{tab:fit_param} as well, and show the corresponding comparisons between model and data in \autoref{fig:perp_fit} and \autoref{fig:drift_fit}. For completeness, we also show our estimate of the parallel diffusion coefficient, computed by multiplying our fits for $(\kappa_\parallel/\kappa_\perp)$ and $\kappa_\perp$, in \autoref{fig:par_fit}. In all cases, we see that the model fits the data reasonably well.

\subsection{Superdiffusion and its implications}
\label{sec:super}

\subsubsection{The ubiquity and origin of superdiffusion}

As discussed \autoref{sec:alpha}, we find that CR transport in our simulations is better characterised by superdiffusion, with an index $\alpha < 2$, than by classical diffusion, $\alpha=2$. Viewed in terms of the Green's function describing the instantaneous distribution of CRs injected at a particular place and time, superdiffusion yields a shape that has a narrower core but more extended wings than a Gaussian, with the amount the wings are extended, and conversely the core is thinned, controlled by $\alpha$. Perhaps more importantly, the rate at which the width of the Green's function expands over time is different for superdiffusion than classical diffusion: the characteristic width of the distribution $\Delta x \propto t^{1/\alpha}$, so for $\alpha<2$ the width $\Delta x$ increases with time more steeply than for classical diffusion.

We see from \autoref{fig:a_Alfven} that $\alpha$ varies little with  $\Mao$ or $\chi$ for either the parallel or perpendicular case, and for most cases the simulations lie in the range $1.4 \lesssim \alpha \lesssim 1.8$. This finding is consistent with studies of CR diffusion on much smaller scales that also find the transport to be superdiffusive \citep{xu2013cosmic,lazarian2014superdiffusion,litvinenko2014analytical}.\footnote{A word of caution is that these authors did not determine the generalised diffusion coefficient rigorously as we do here, because they compute their diffusion coefficients directly from the time averaged-spatial variance of the CR particles, as $\kappa \sim \langle (\Delta x)^2 \rangle  / 2t$. In superdiffusion the time rate of increase of $\langle (\Delta x)^2 \rangle$ is non-linear, so this method is not appropriate, and does not yield an accurate estimate of the diffusion coefficient when $\alpha \neq 2$.} The typical index of $\alpha\approx 1.5$ that we measure can plausibly be explained as a result of Richardson diffusion, whereby magnetic field lines (or any other quantity advected with a turbulent flow) diverge at a rate of $\langle |x_1(t) - x_2(t) |^2 \rangle \sim t^{3}$ \citep{kupiainen2003nondeterministic,lazarian2012relation}, which corresponds to $\alpha=3/2$.

\subsubsection{Observational implications of superdiffusion}
\label{sssec:obs_super}

\begin{figure}
    \centering
    \includegraphics[width=0.47\textwidth]{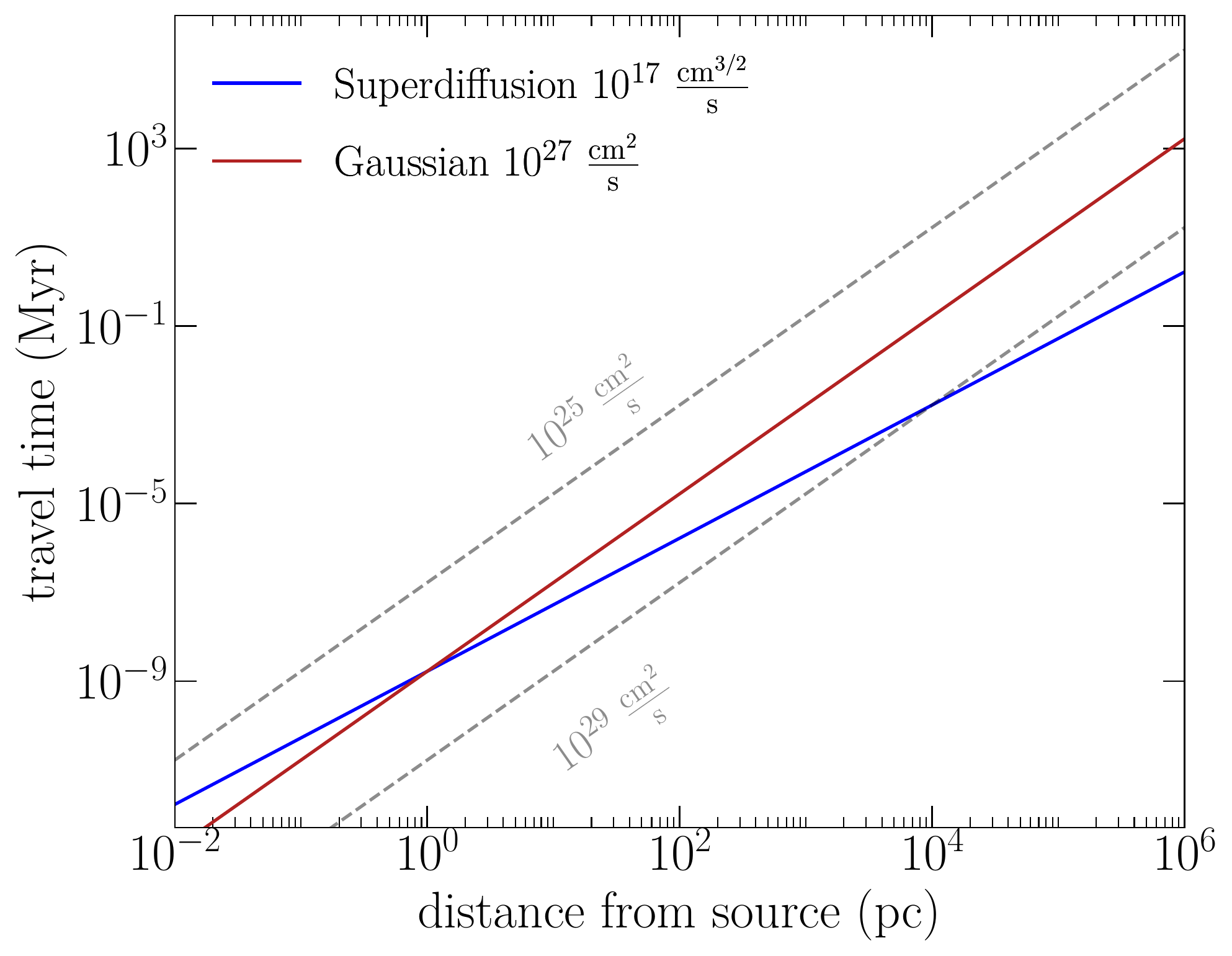}
    \caption{Schematic diagram of cosmic ray travel times $t = 2 \ell^{\alpha}/\kappa$ as a function of distance from source location. The red line indicates the expected travel time assuming a Gaussian model ($\alpha=2$) with $\kappa = 10^{27}$ cm$^2$ s$^{-1}$ (with the dashed lines showing a factor of 100 larger and smaller coefficient for reference), with the blue line showing the expected times for a superdiffusive ($\alpha = 3/2$, $\kappa = 10^{17}$ cm$^{3/2}$ s$^{-1}$) diffusion model.}
    \label{fig:arrival_times}
\end{figure}

Our finding that CR transport at galactic scales is likely superdiffusive rather than classically diffusive has important implications for the interpretation of observations. To understand these implications, we begin by pointing out that some care is required to compare the results we obtain with diffusion coefficients reported in the literature, as these quantities do not have the same units. For example,  a diffusion coefficient corresponding to an index $\alpha = 1.5$, has units of $\text{length}^{1.5} / \text{time}$. Clearly, this is not directly comparable to a classical diffusion coefficient with units of length squared per time. The two can only be compared at a particular, specified length scale, by considering the characteristic time required to travel that distance from a source. That is, the characteristic time required for a classical diffusive process with diffusion coefficient $\kappa_{\rm class}$ to transport CRs a distance $\ell$ is $t_{\rm class} \sim 2\ell^2/\kappa_{\rm class}$, while the time scale required for a superdiffusive process of index $\alpha$ and coefficient $\kappa_{\rm super}$ to transport CRs the same distance is $t_{\rm super} \sim 2\ell^{\alpha}/\kappa_{\rm super}$; the ratio of these two times is $t_{\rm class}/t_{\rm super} =  \ell^{2-\alpha} (\kappa_{\rm super}/\kappa_{\rm class})$. The fact that this ratio depends on $\ell$ means that one cannot meaningfully ask whether superdiffusion is faster or slower than classical diffusion; the answer depends on the length scale over which they are being compared. We illustrate this in \autoref{fig:arrival_times}, where we show characteristic travel times $t$ as a function of length scale $\ell$ for example classical diffusion and superdiffusion coefficients. In the example shown, superdiffusive transport with $\kappa_{\rm super} = 10^{17}$ cm$^{3/2}$ s$^{-1}$ yields a travel time comparable to classical diffusion with a coefficient $\kappa_{\rm class} = 10^{27}$ cm$^2$ s$^{-1}$ for $\ell\sim 3$ pc, but more closely resembles classical diffusion with $\kappa_{\rm class} = 10^{29}$ cm$^2$ s$^{-1}$ for $\ell \sim 300$ pc.

This matters for the interpretation of observations because most observational methods of diagnosing diffusion coefficients are ultimately sensitive to CR travel times, which are then converted to diffusion coefficients assuming that diffusion is the correct description of CR transport. For example, the grammage through which a CR population passes, as deduced from its boron (B) to carbon (C) ratio \citep[e.g.,][]{adriani2014measurement,genolini2015theoretical,evoli2020ams}, is simply the product of the ISM density the CRs encounter, the speed of light, and the time taken by the population to reach the observer. Similarly, measurements of changes in radio synchrotron or $\gamma$-ray spectral index with distance off a galactic plane \citep[e.g.,][]{bloemen1993galactic,castellina2005diffusion,evoli2008cosmic,gabici2010constraints,genolini2015theoretical,lopez2022gamma} are mainly sensitive to the amount of time for which the CRs producing the emission are subject to loss processes (pion production, inverse Compton and synchrotron radiation) before reaching a given distance off the plane. If such an observation is interpreted in terms of a classical diffusion model, but CR transport is actually superdiffusive, then the classical diffusion coefficient that one deduces will depend on the length scales probed by the measurement. Again examining \autoref{fig:arrival_times}, we can see that superdiffusive transport with a fixed coefficient will appear, if interpreted assuming a classical diffusion model, as diffusion with a larger coefficient on larger scales and a smaller coefficient on smaller scales. Significantly, while there are a huge diversity of results for CR diffusion coefficients available in the literature, there does appear to be tension where studies that infer their diffusion coefficient from B/C ratios \citep{1983A&A...125..249L}, which are mostly sensitive to CR sources relatively close to the Earth and for which the CRs have travelled $\ll 1$ kpc, tend to yield lower diffusion coefficients than those based on fitting the $\gamma$-ray or radio synchrotron distribution at $\gtrsim$kpc distances off the galactic plane, which generally favour larger coefficients. If CR transport is superdiffusive rather than classically diffusive, it  
may alleviate this apparent tension.
%possible that these results are in fact compatible.

%%%%%%%%%%%%%%%%%%%%%%%%%%%%%%%%%%%%%%%%%%%%%%%%%%%%%%%%%%%%%%%%%%%%%
%%% Astrophysical Context of Results
%%%%%%%%%%%%%%%%%%%%%%%%%%%%%%%%%%%%%%%%%%%%%%%%%%%%%%%%%%%%%%%%%%%%%
\subsection{Comparison to previous results}
\label{sec:context}
When comparing our results to previous work, it is important to reiterate the distinction between the scales of CR transport studies. There has been limited work on the scaling of macro-physical diffusion (i.e., on scales comparable to the background flow correlation scale) with $\M,\,\Mao$ and $\chi$. Hence, to compare our results with the literature, we look at \textit{micro}-physical simulations (on scales comparable to the CR scattering mean free path) first, pointing out the major differences that will affect the diffusion coefficient calculations. In the latter, the gyroradius is resolved and CRs travel along a field line at $\approx c$, compared to $\vai$ in our simulations. Therefore there will be no dependence on $\chi$. A further implication of taking a micro-physical approach  is that since the CR speed, $c$, is much greater than the gas velocity, the dynamics of the turbulence will evolve on significantly different timescales to CR transport. Effectively, one may consider the micro-physical studies to be performing frozen time simulations on the MHD turbulence \citep[such as:][]{yan2008cosmic,xu2013cosmic}. Thus, one of our main diffusive mechanisms, field line tangling and stretching, is attenuated (since the field is frozen rather than time-dependent), and another, field line advection, is absent entirely (see \autoref{sec:random_walk} for more on field line advection). With these caveats in mind, we may still make useful comparisons about the general picture of CR diffusion found here and in the prior literature.

One of our significant results is that $\kappa_{\perp} = \kappa_{\parallel}$ for $\Mao \gtrsim 2$. This is consistent with earlier theoretical studies, which predict that CR diffusion is isotropic when measured from scales larger than the coherence length of the $\mathbf{B}$-field lines \citep{casse2001transport,yan2008cosmic,Krumholz2020CosmicGalaxies}. \citet{casse2001transport} quantify the anisotropy as $ \kappa_{\perp} = \eta^{2.3 \pm 0.2}\kappa_{\parallel}$, where $\eta$, which has  a maximum of 1, characterises the strength of the turbulence.  This is generally in agreement with our results from \autoref{fig:iso} in which $\kappa_{\parallel} / \kappa_{\perp} \to 1$ as $\Mao \to 2$. However, there is less studies, and less agreement in the literature regarding the behaviour of $\kappa_\parallel$ with $\Mao$ in the $\Mao\lesssim 1$ regime. Here a number of authors report $\kappa_\parallel \sim \mathcal{M}_A^4$ \citep{yan2008cosmic,xu2013cosmic,cohet2016cosmic}\footnote{Note that our description of this result in terms of the total Alfv\'en Mach number $\mathcal{M}_A$, rather than the large-scale field Alfv\'en Mach number $\Mao$ that we use, is intentional. The distinction is that, in simulations such as ours where the field is self-consistently evolved to steady state rather than frozen in time, it is possible to have $\Mao \gg \mathcal{M}_A$ as a result of dynamo amplification, a process that obviously does not occur in simulations using frozen fields. However, even in fully time-dependent simulations such as ours the difference between $\Mao$ and $\mathcal{M}_A$ is significant only when $\Mao \gtrsim 1$, and thus the distinction is small in the regime we are currently discussing.}, i.e., the diffusion scaling exactly with the magnetic field variance \citep{Beattie2022_va_fluctuations}. While the direction of the scaling of $\kappa_{\parallel}$ for sub-Alfv\'enic turbulence found in these simulations is qualitatively consistent with our findings, we find a vastly different scaling, $\kappa_{\parallel} \sim \Mao^{-1}$, which is closer to results from \citet{casse2001transport}. This difference may well be due to the physical differences in scale discussed above, in that the dominant diffusion mechanism in the perpendicular direction when $\Mao \ll 1$ -- field line advection -- only occurs in larger-scale simulations such as ours, and is necessarily absent in smaller-scale frozen-field simulations.

While most previous numerical studies of CR transport have focused on the micro- and mesoscopic scales, both observations and simulations on galactic or cosmological scales are sensitive almost exclusively to transport on much larger scales. We may use our results to make some sample calculations about macroscopic CR diffusion rates expected in the ISM. A typical region of the warm atomic ISM, of the type that fills most of the volume of the thin disc, is characterised by 
a total velocity dispersion $\sigma_V = 8$ km s$^{-1}$, ionisation fraction $\chi=10^{-2}$, outer scale of turbulence comparable to the twice the warm gas galactic scale height, $\ell_0 \approx 300$ pc, and Alfv\'en Mach number $\Mao\approx 2$ \citep[e.g.,][]{Wolfire03a}. The warm ionised phase, which fills the thick gaseous disc, has similar $\Mao$, but $\chi\approx 1$, $\ell_0\approx 3$ kpc, and $\sigma_V\approx 20$ km s$^{-1}$ \citep{boulares1990galactic}. Inserting these figures into our fitting formula, \autoref{eq:fit_func}, using our best-fit parameters from \autoref{tab:fit_param}, gives $\kappa_\parallel/\kappa_\perp \approx 1$ in both cases (i.e., transport is close to isotropic), and $\kappa_\parallel \approx \kappa_\perp \approx 0.5\ell_0^\alpha \tau^{-1} = 1.2 \times 10^{16}$ cm$^{3/2}$ s$^{-1}$ for the thin atomic disc and $\kappa_\parallel \approx \kappa_\perp \approx 0.14\ell_0^\alpha \tau^{-1} = 3 \times 10^{16}$ cm$^{3/2}$ s$^{-1}$ for the thick ionised disc; in both cases we have adopted $\alpha=3/2$ for the numerical evaluation. Recalling our earlier discussion in \autoref{sssec:obs_super}, superdiffusive transport with the coefficient we have estimated for the thin neutral disc would, if interpreted assuming classically diffusive transport, appear to correspond to a diffusion coefficient of $\kappa_{\rm class} \approx 4\times 10^{26}$ cm$^2$ s$^{-1}$ on 300 pc scales (about the outer scale of WNM turbulence). Using our parameters for the thicker ionised disc, we would infer a diffusion coefficient of $3\times 10^{27}$ cm$^2$ s$^{-1}$ on 3 kpc scales (about the characteristic width of the synchrotron-emitting regions seen around Milky Way-like galaxies; \citealt{Krause18a}). These values are similar to values found by \citet{xu2022cosmic} of $\sim 10^{26} - 10^{28} \ \rm{cm}^2\rm{s}^{-1}$ who used methods suggested by \citet{Krumholz2020CosmicGalaxies} to calculate macroscopic diffusion coefficients for $\sim \ 1 - 100$ GeV streaming CRs.

%an outer scale of turbulence  of $\ell_0 = 100\,\text{pc}$, local sound speed $c_s = 2 \, \rm{kms^{-1}}$, $\Mao \approx 2$, $\mathcal{M} \approx 4$, $\chi = 10^{-2}$ and we show an $\alpha = 3/2$ and 2.
%Then we expect
%\begin{align}
%\label{eqn:sample}
%     \kappa_{\rm{iso}} &\approx \frac{0.05}{2.25\chi + 0.1} \left(\tanh(\Mao - 2) + (2.25\chi + 1.1) \right) \  \ell_0^{\alpha} \tau^{-1} \\ \nonumber
%     &\approx \frac{0.05}{2.25 \times 10^{-2}+ 0.1} \left((2.25 \times 10^{-2} + 1.1) \right) \  \ell_0^{\alpha} \tau^{-1} \\\nonumber
%     &\approx 0.45 \ell_0^{\alpha} \ \tau^{-1} \\\nonumber
%     &\approx 1 \times 10^{26} \ \rm{cm}^2 \rm{s}^{-1} \ \ \ \ \ \ (\text{for $\alpha = 2$})\\\nonumber
%    &\approx 6 \times 10^{15} \ \rm{cm}^{3/2} \rm{s}^{-1} \ \ \ (\text{for $\alpha = 3/2$}) \nonumber
%\end{align}

%\begin{align}
%\label{eqn:sample}
%    \kappa &\approx \frac{\ell_0 \sigma_V}{\Mao \sqrt{\chi}} \\\nonumber &\approx 1.2 \times 10^{27} \  \left( \frac{\sigma_V}{8 \ \rm{km}\rm{s^{-1}}} \right)  \left( \frac{\ell_0}{100 \ \rm{pc}} \right) \left( \frac{2}{\Mao} \right) \left( \frac{10^{-2}}{\chi} \right)^{1/2} \, \text{cm}^2\text{s}^{-1}.
%\end{align}
%The value calculated in \autoref{eqn:sample} are on the lower end of the spectrum of empirically measured CR diffusion coefficients (see \autoref{tbl:values}). 

%%%%%%%%%%%%%%%%%%%%%%%%%%%%%%%%%%%%%%%%%%%%%%%%%%%%%%%%%%%%%%%%%%%%%
%%% Limitations
%%%%%%%%%%%%%%%%%%%%%%%%%%%%%%%%%%%%%%%%%%%%%%%%%%%%%%%%%%%%%%%%%%%%%
\subsection{Limitations of study}
\label{sec:Limitation}
Here, we outline some key limitations of this study. Some of these are specific to the work undertaken in this study and reflect the parameter space that has been explored (\autoref{sec:sub_sonic}), while others represent fundamental limitations of the physical assumptions we make when carrying out our simulations (\autoref{sec:back_react}).

\subsubsection{Parameter space explored}
\label{sec:sub_sonic}
In all of our trials, we have assumed an isothermal ISM with no phase structure. In reality, the ISM is a multiphase plasma \citep{diamond1989structure,spaans1997star,hennebelle2008warm,krumholz2012star,gent2013supernova, mandal2020molecular,Seta2022}. While each individual phase is approximately isothermal, CRs may not be confined to regions dominated by a single phase. Therefore, we have not explored how CR propagation is modified when CRs are able to cross from one phase to another, in which the variation in $\chi$ between phases would be significant. 

%We also have assumed a constant $\chi$ for the entire plasma, whereas the ionisation fraction is likely to be density-dependent in reality. Although the extent to which the ionisation state equilibrates in short-lived density fluctuations is an open question. The extent to which the ionisation fractions respond to changes in density would limit the effect density fluctuations have on macroscopic diffusion. This is because the ionisation fraction decreases with increasing density in equilibrium to first order as $\chi\propto n^{-1/2}$, so fluctuations in the ion density are smaller than fluctuations in the total density.

A second limitation is that we have not explored sub-sonic turbulence ($\M < 1$), which differs from the supersonic regime we have explored due to the lack of density variations and compressible modes and dominance of Alfv\'en modes \citep{kowal2007scaling,kowal2007density,esquivel2010tsallis,burkhart2009characterizing,beattie2021multishock}. While observations show that the cooler phases of the ISM are highly supersonic, warmer phases such as the warm ionised and warm neutral medium are only mildly supersonic ($\mathcal{M} \sim 1$), and the intracluster medium is subsonic \citep{cho2019mhd}. As discussed in \autoref{sec:dens_fluc}, density fluctuations are the main driver of parallel CR diffusion in the sub-Alfv\'enic regime. This suggests that CR transport for subsonic, sub-Alfv\'enic turbulence may be fundamentally different from the supersonic, sub-Alfv\'enic regime we have explored thus far. Conversely, however, since density fluctuations are less important in super-Alfv\'enic transport, extending to the subsonic regime would likely have little effect on our super-Alfv\'enic results.

Another limitation, resulting from the restricted parameter space used in this study, is that we have a driving scale of $\ell_0 = L/2$ in all of our MHD simulations. %This means we are always simulating eight dominant ``eddies" (or eight statistically independent volumes in the turbulent velocity). 
We have not explored the effects of reducing the driving scale, and whether we may see different behaviour in $\kappa$ as $\Mao \ll 1$ in this case. Somewhat connected to this limitation is the restricted range of the turbulence cascade $\ell_0 \ll \ell \ll \ell_\nu$, where $\ell_\nu$ is the numerical viscous dissipation scale, in the MHD simulations with grid resolutions of $576^3$ \citep{kitsionas2009,federrath2010comparing,federrath2013}. This limited resolution may affect the values of both the diffusion coefficients and the superdiffusion index parameter $\alpha$. We discuss convergence tests in \aref{sec:converge}. The dynamic range of the turbulent cascade increases linearly with grid resolution \citep[Appendix C in ][]{mckee2020}; hence we probe the change in size of the cascade by roughly an order of magnitude between grid resolutions $\sim 100^3$ and $1,000^3$. Over this range, we do not find any systematic trends between the grid resolution and either of these parameter values. Thus, resolving high-$k$ modes in the turbulence cascade does not appear to be critical to the results concerning large-scale CR diffusion coefficients or the nature of the diffusion reported in this study.

A final limitation inherited from our MHD data set comes from our choice to drive all simulations using a ``natural mixture" of compressive and solenoidal modes. As driving becomes more compressive, density fluctuations increase and large voids and filamentary structures dominate the density field \citep{federrath2010comparing,cohet2016cosmic,jin2017effective}. This in turn will affect parallel CR diffusion in the anisotropic regime, where density fluctuations are a dominant generator of dispersion, \citep[see][for more detail]{Beattie2022_va_fluctuations}. Consistent with this general idea, \citet{cohet2016cosmic} find that compressively- and solenoidally-driven turbulence produce different levels of CR diffusion, but the differences vanish as $\Mao \to 1$. Mapping out the full effects of the driving modes in the anisotropic regime will require a larger parameter study than either \citet{cohet2016cosmic} or we have performed.

\subsubsection{Physical assumptions}
\label{sec:back_react}
We briefly reviewed our treatment of SCR transport in \autoref{sec:criptic}.
This model has certain limitations. One is that we assume \textit{one-way} interaction of the plasma with the SCRs, i.e., we allow the plasma to affect the propagation of SCRs, but we neglect the effects of SCR forces on the gas. 
%This limitation is imposed by the fact that our pipeline consists of a first step using an MHD simulation code (\textsc{flash}) and a second step using \textsc{Criptic}, a post-processing tool that calculates the transport of CRs through a pre-defined plasma background. Thus, there is no way for SCR packets in \textsc{Criptic} to influence the evolution of the background plasma, and no energy transfers from the SCRs to the plasma. 
To the extent that such back-reactions themselves modify CR transport (as they must when the CR pressure gradient becomes high enough to modify the statistical properties of the turbulence), our simulations will not capture those effects. The observed CR pressure in the Milky Way is comparable to the ram pressure on large scales \citep{spitzer1978physical,boulares1990galactic,ferriere2001interstellar,Grenier2015}, but the CR distribution is much smoother than the gas distribution. This means that CR forces are likely not dominant at the galactic midplane, suggesting that our calculation is reasonable. However, this assumption is perhaps more questionable in regions of strong galactic wind or near local strong sources of CRs, and it may also be problematic for star-forming dwarf galaxies \citep{Crocker2020CosmicCalorimetryb}. It should be pointed out, however, that a self-consistent MHD plus CR fluid code would do no better in this circumstance, unless the code reached resolutions sufficient to capture the full MHD turbulent cascade responsible for diffusion of CRs via the mechanisms we have identified.

A second limitation is that our CR transport model is not applicable to cosmic rays with energies $\gtrsim 0.1 - 10$ TeV. We assume that the CR pitch angle distribution is isotropised on scales smaller than those being simulated, so the distribution of SCRs is adequately described by the pitch-angle-averaged Fokker-Planck equation. As CR energy increases the rate of pitch angle scattering via the streaming instability decreases (due to the decreasing density of CRs energetic enough to drive resonant Alfv\'en waves), both of which contribute to an increase in the CR isotropisation scale. \citet{Krumholz2020CosmicGalaxies} estimate that our basic assumption -- that CRs stream at the ion Alfv\'en speed -- will begin to fail when the streaming instability becomes too weak for CR energies above $\approx 0.1 - 10$ TeV. Where $\sim 0.1$ TeV being more relevant to normal star-forming galaxies like the Milky Way, and $\sim 10$ TeV to denser, more strongly magnetised starbursts. In either case our treatment is valid for the range of CR energies that is most important for affecting the dynamics and chemistry of the ISM \citep{padovani2020impact}.

\section{Conclusions}
\label{Conclusions}
%%%%%%%%%%%%%%%%%%%%%%%%%%%%%%%%%%%%%%%%%%%%%%%%%%%%%%%%%%%%%%
%%%%%%%%% Conclusions
%%%%%%%%%%%%%%%%%%%%%%%%%%%%%%%%%%%%%%%%%%%%%%%%%%%%%%%%%%%%%%
In this study, we present results from simulations of streaming CRs (SCRs) through a supersonic, magnetized medium, with the aim of developing an effective theory that describes the transport of such SCRs as a function of the plasma Mach number $\mathcal{M}$, the mean-field Alfv\'en Mach number $\Mao$, and the ionisation fraction $\chi$. We use an extensive library of MHD and CR propagation simulations to explore this parameter space, and use the simulations to inform a turbulent transport model. The theory we develop is suitable for use in analytic calculations or simulations where turbulent structures are unresolved, and has important implications for the interpretation of observations used to infer SCR diffusion coefficients. Our main results are summarised below.

\begin{itemize}
    \item We identify two distinct regimes of SCR diffusion. When the Alfv\'en Mach number of the turbulence $\Mao \lesssim 0.5$, we have an anisotropic regime where parallel diffusion is substantially more rapid than perpendicular diffusion, $\kappa_{\parallel} > \kappa_{\perp}$. In this regime, the parallel diffusion is dominated by density fluctuations that randomly modulate the streaming speed, causing SCRs traveling along parallel field lines to diverge. The parallel diffusion rate is therefore primarily governed by $\chi$, which sets the mean streaming speed, with a lesser dependence on $\M$ due to the effect of $\M$ on the  strength of the density fluctuations. Perpendicular diffusion in this regime is caused by field line advection by turbulent flows, leading to a diffusion rate that, measured relative to the turbulent turnover time, is nearly constant aside from a weak dependence on $\M$.\\
    \item For $\Mao \gtrsim 2$, by contrast, we find an isotropic regime where $\kappa_{\parallel} = \kappa_{\perp}$, and diffusion is dominated by the tangling and turbulent transport of field lines as CRs stream down them. The diffusion rate is sensitive to the streaming speed, and thus to $\chi$, when $\chi$ is not too small compared to unity, but when streaming is very fast it becomes limited by the time required for the cosmic rays to travel through space-filling, highly-tangled $\mathbf{B}$ field lines. Our finding that diffusion is approximately isotropic when $\Mao\gtrsim 2$ provides a physical explanation for why models that assume a single isotropic diffusion coefficient appear to provide a good match to large-scale observations of galaxies.\\
    \item We find that almost ubiquitously in our parameter space SCR transport is better described by superdiffusion with an index $\alpha\approx 1.5$, rather than classical diffusion, $\alpha = 2$. We propose that this may explain the apparent discrepancies between observationally-inferred CR diffusion coefficients that are effectively measured at different length scales: if one interprets superdiffusive transport in terms of classical diffusion, and attempts to assign a diffusion coefficient, it will appear that the diffusion coefficient is larger on larger scales and smaller on smaller scales, exactly as many studies of CR diffusion seem to report.\\
    \item  We provide fitting formulae for the effective diffusion coefficients and streaming speeds of SCRs subject to unresolved MHD turbulence, calibrated based on our simulation library. These provide sub-grid recipes suited to the inclusion of physically motivated SCR diffusion prescriptions in simulations of galaxy evolution.
\end{itemize} 
This work provides the crucial first steps in bridging the gap between the disparate scales of CR transport and cosmological simulations, increasing physical accuracy of simulations and allowing for a deeper understanding of the evolution of galaxies.

\section*{Acknowledgements}
We thank the anonymous reviewer for providing constructive feedback of our study. M. L.~S. thanks Angelina Yan for insightful comments and support throughout study and acknowledges financial support from the Australian National University and the Australian Government via the Australian Government Research Training Program Fee-Offset Scholarship.

J.~R.~B.~thanks Christoph Federrath's and Mark Krumholz's research groups for many productive discussions and acknowledges financial support from the Australian National University, via the Deakin PhD and Dean's Higher Degree Research (theoretical physics) Scholarships and the Australian Government via the Australian Government Research Training Program Fee-Offset Scholarship. 

M.~R.~K. and R.~M.~C. acknowledge support from the Australian Research Council through its \textit{Discovery Projects} scheme, award DP190101258.

C.~F.~acknowledges funding provided by the Australian Research Council (Future Fellowship FT180100495), and the Australia-Germany Joint Research Cooperation Scheme (UA-DAAD). 

We further acknowledge high-performance computing resources provided by the Australian National Computational Infrastructure (grant~ek9 and~jh2) in the framework of the National Computational Merit Allocation Scheme and the ANU Merit Allocation Scheme, and by the Leibniz Rechenzentrum and the Gauss Centre for Supercomputing (grant~pr32lo and~pn73fi). 
%The simulation software FLASH was in part developed by the DOE-supported Flash Center for Computational Science at the University of Chicago.

The fluid simulation software, \textsc{flash}, was in part developed by the Flash Centre for Computational Science at the University of Chicago and the University of Rochester. The cosmic ray simulation software \textsc{criptic} was developed by Krumholz et al.~2022 (in prep). Data analysis and visualisation software used in this study: \textsc{C++} \citep{Stroustrup2013}, \textsc{numpy} \citep{Oliphant2006,numpy2020}, \textsc{matplotlib} \citep{Hunter2007}, \textsc{scipy} \citep{Virtanen2020}, \textsc{pylevy} \citep{pylevy} and \textsc{emcee} \citep{foreman2013emcee}.

%%%%%%%%%%%%%%%%%%%%%%%%%%%%%%%%%%%%%%%%%%%%%%%%%%
\section*{Data Availability}
The data underlying this article will be shared on reasonable request to the corresponding author.



\bibliographystyle{mnras}
\bibliography{Refs} % if your bibtex file is called example.bib

%%% APPENDIX %%%
\appendix
\section{Convergence testing}
\label{sec:converge}

As discussed in \autoref{Methods}, our simulation pipeline requires a number of choices regarding the initial numerical configurations of either the MHD fluid simulations, or the particle simulations. Here we summarise a series of tests we carried out to determine the spatial, temporal, and particle resolution needed for the study. For each of these of these tests we assess convergence by running the full simulation pipeline, fixing all parameters except the one we are testing (i.e. a grid resolution $576^3$, 10 MHD realisations per $\tau$, an injection rate of $n_{\rm{CR}} / \tau = 10^6$ and $9^2$ sources), and deriving parallel and perpendicular diffusion coefficients as a function of the resolution parameter being studied. We carry out this test using trials \texttt{M4MA10C0} and \texttt{M4MA10C4} to sample the high $\Mao$ regime over a wide range in $\chi$, and either \texttt{M4MA05C0}, and \texttt{M4MA05C4} or \texttt{M4MA01C0}, and \texttt{M4MA01C4} to amply sample the low $\Mao$ regime.

\subsection{Spatial resolution}
\begin{figure*}
    \centering
    \includegraphics[width=\textwidth]{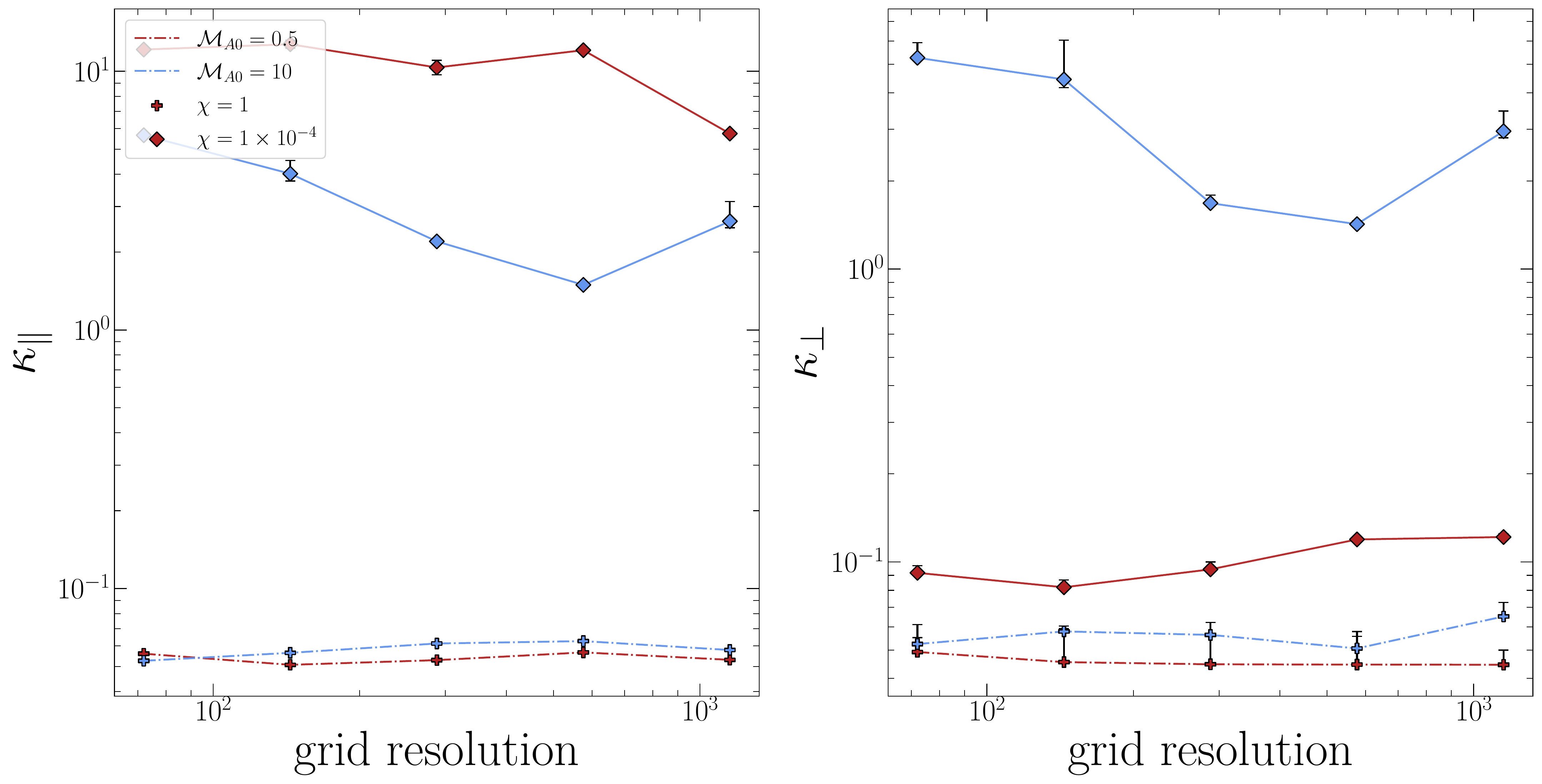}
    \caption{Results of grid resolution convergence test, where we show trials for $\Mao = 0.5$ (red) and $\Mao = 10$ (blue), with $\chi = 1$ and $1 \times 10^{-4}$ which are indicated by the marker types. We see strong convergence in almost all cases at resolutions of $576^3$. }
    \label{fig:grid_c}
\end{figure*}

\begin{figure*}
    \centering
    \includegraphics[width=\textwidth]{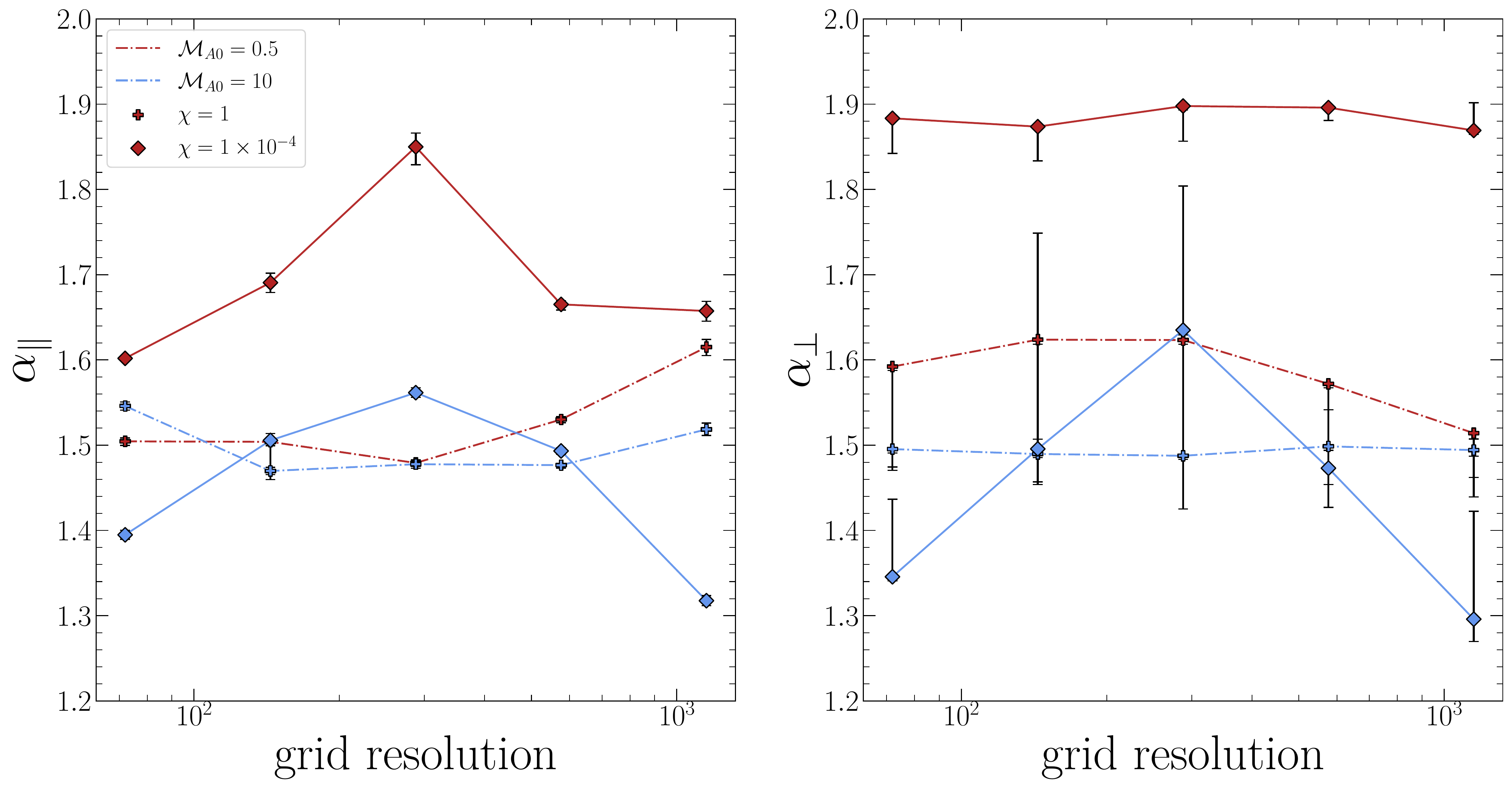}
    \caption{Similar to \autoref{fig:grid_c}, but for super-diffusion parameter $\alpha$ as a function of numerical grid resolution.
    We find that both the parallel and perpendicular super-diffusion parameters show no systematic variation
    with grid resolution.}
    \label{fig:grid_c_a}
\end{figure*}
To test for convergence in grid resolution we systematically vary the number of grid cells used for the discretisation of plasma simulations between $72^3 - 1152^3$, increasing by factors of 2. We show the results of this test in \autoref{fig:grid_c} and \autoref{fig:grid_c_a}, which show the fitted diffusion coefficient $\kappa$ and super-diffusion parameter $\alpha$, respectively. The experiments that vary the most with resolution are those with high $\Mao$ and low $\chi$, which show substantial resolution dependence at lower grid resolutions tending to have reduced diffusion at higher resolution; only the $\chi=10^{-4}$, $\Mao=10$ case shows noticeable differences between resolutions of $576^3$ and $1152^3$, and even these are at the $\approx 30\%$ level. All other cases are extremely well converged at $576^3$. 
%While further increasing the grid resolution might improve the results further, it is not computationally feasible for us to do so, and any remaining resolution effects are likely to be smaller than the uncertainties inherent in our adoption of a parametric fitting model.

The relative insensitivity of our results to resolution might at first seem surprising, given that other authors studying cosmic ray diffusion through turbulence have found that the results are sensitive to resolution \citep[e.g.,][]{cohet2016cosmic}. However, it is important to recall that, because we are studying $\sim$GeV cosmic rays whose pitch angles isotropise on unresolved scales, and thus whose effective velocity is $\ll c$, our diffusion mechanisms are very different than those studied by earlier authors, who have focused on much higher energy cosmic rays whose gyroradii are resolved by the simulation, and that move through the background plasma at $c$. In the latter case, the field is effectively frozen (since the cosmic ray speed is vastly greater than the plasma speed), and the main diffusion mechanism is via resonances between structures in the frozen magnetic field and the gyroscopic motion. Resolution matters for this because the resonant structures are generally small compared to the total simulation volume, and thus high resolution is required to capture the turbulent cascades that create them. This process does not operate in our simulations since we do not follow individual cosmic rays on scales comparable to the gyroradius. Diffusion is instead driven by the mechanisms discussed in \autoref{sssec:mechanisms}. A crucial difference is that all of these processes -- field line tangling and stretching, field line advection, gas flow along the field, and density fluctuations -- are driven primarily by the largest eddies, which contain the bulk of the turbulent power and have the largest associated velocities and magnetic field perturbations. This makes the resolution requirements much less severe, since the majority of the effect is driven by large-scale rather than small-scale structures.

\subsection{Temporal resolution}
\begin{figure*}
    \centering
    \includegraphics[width=\textwidth]{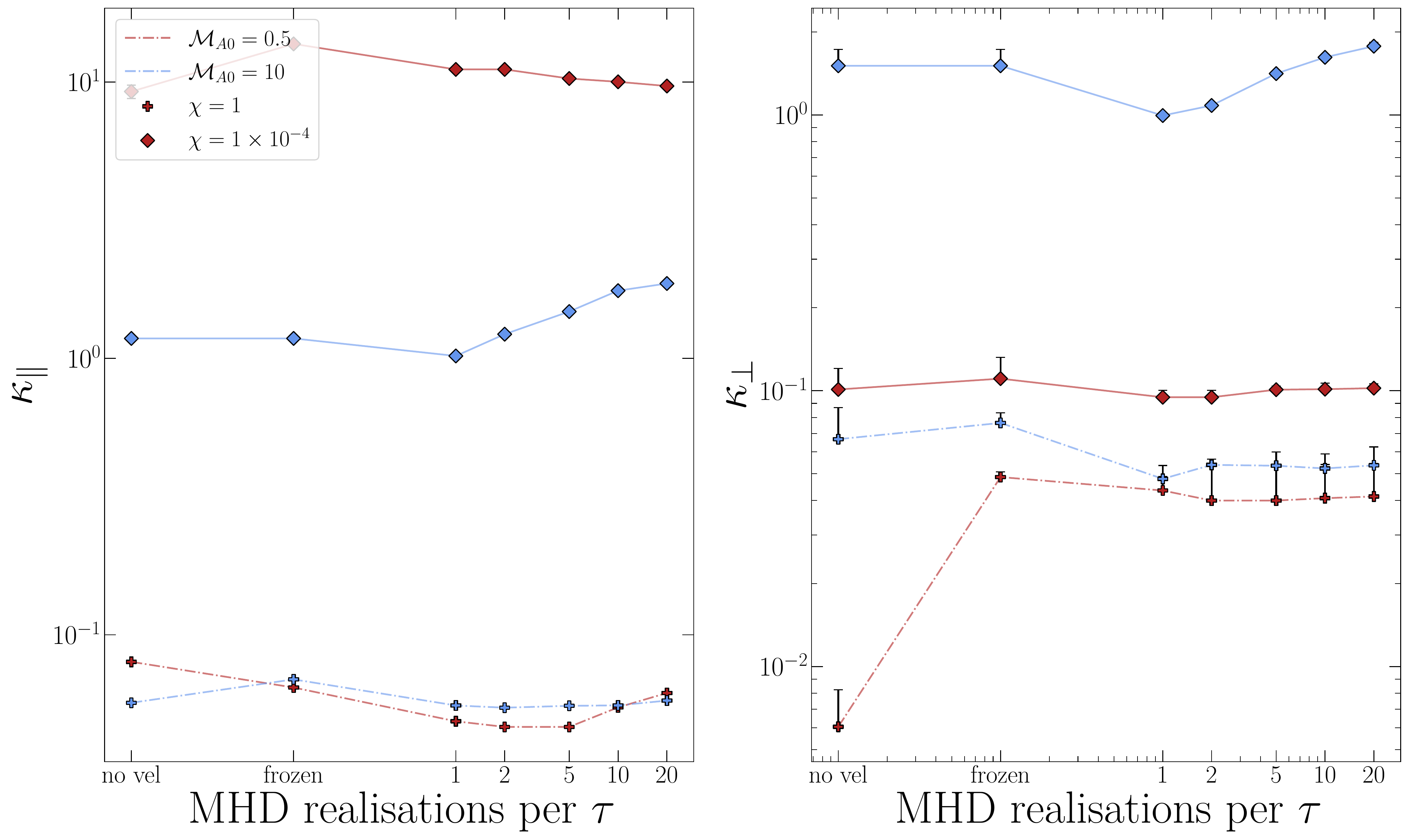}
    \caption{Temporal resolution tests for the same trials as in \autoref{fig:grid_c}. The x-axis is shown in units of MHD realisations per $\tau$, with the \textit{frozen} trials indicating a frozen field (i.e., one that does not evolve in time at all), and the \textit{no vel} trials indicating cases where both the magnetic field is frozen and we set the gas velocity to zero.}
    \label{fig:time_c1}
\end{figure*}

We next check convergence in the time resolution with which we sample MHD realisations of the plasma simulations for use in the \textsc{criptic} CR propagation simulation calculation. We parameterise this in terms of the number of MHD realisations per turbulent turnover time $\tau$, and remind the reader that all of our \textsc{criptic} simulations run for $5\tau$. We test for values from $\tau = 1 - 20$; for comparison, we also test one case in which we simply freeze the MHD field but, non-self-consistently, continue to advect the CR packets using the gas velocity field (the ``frozen'' case), and another test where we both use only a single, frozen magnetic field structure and also set the velocities to zero (the ``no vel'' case). From \autoref{fig:time_c1}, we see increasing the temporal resolution has little impact once at least 10 realisations is used per $\tau$, hence we use 10 realisations per $\tau$ for our study. Surprisingly, diffusion coefficients do not change dramatically even if we use a frozen field, and only the $\chi = 1$, $\Mao=0.5$ case changes dramatically if we disable velocity advection (which has the effect of turning off field line advection, the dominant diffusion mechanism in the perpendicular direction for this plasma regime). However, this result can be understood if we recall that we are in this test still using a grid of 81 different CR injection sites, each of which is sampling a spatially distinct part of the magnetic field structure. The fact that the time evolution of the field is relatively unimportant can therefore be seen as a manifestation of the ergodic nature of turbulence, i.e., $\Exp{\mathbf{X}(\mathbf{r},t)}_{t} \sim \Exp{\mathbf{X}(\mathbf{r},t)}_{\mathbf{r}}$, for statistical quantity $\mathbf{X}(\mathbf{r},t)$.

\subsection{Particle resolution}
\begin{figure*}
    \includegraphics[width=0.95\textwidth]{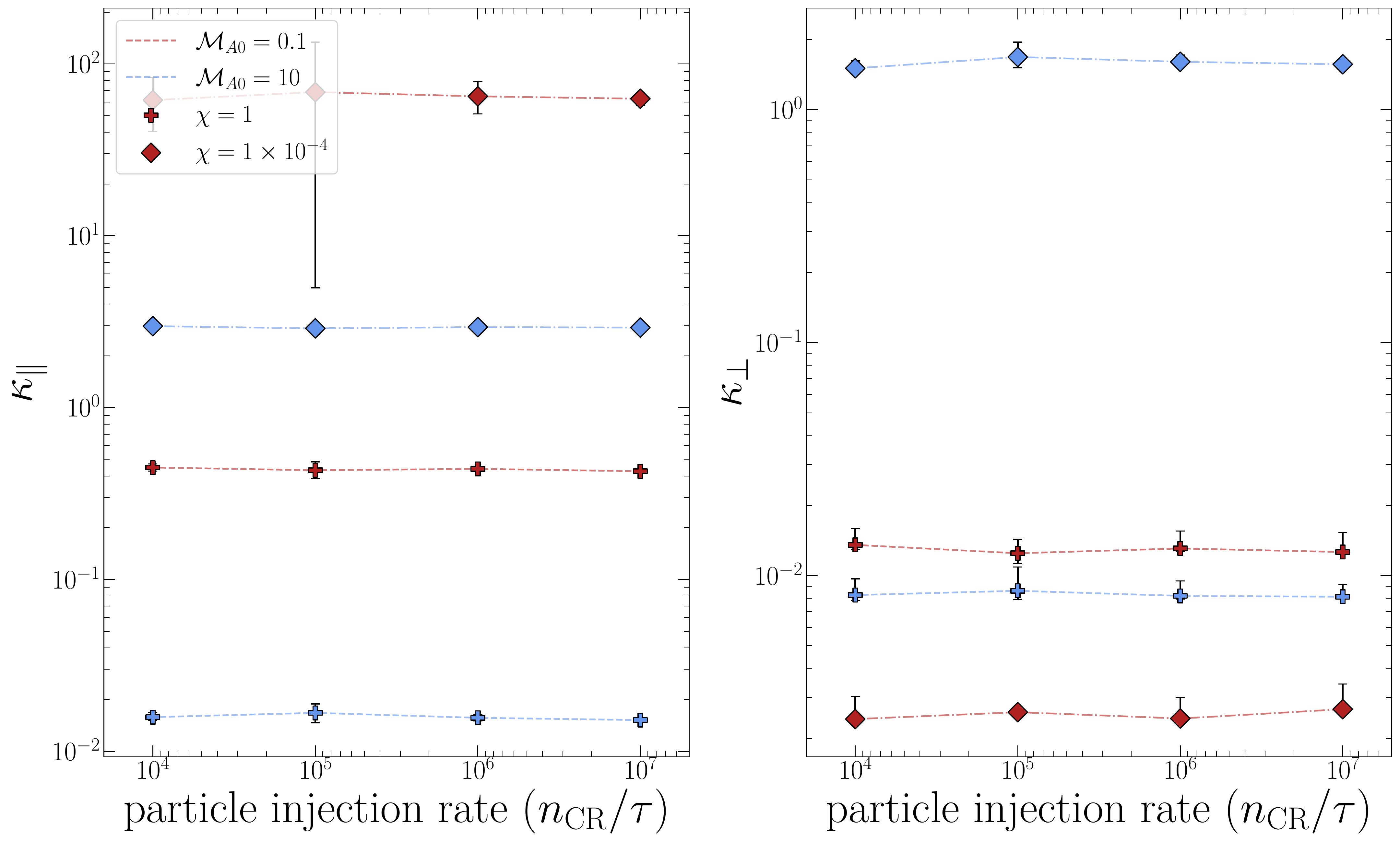}
    \caption{Similar to \autoref{fig:time_c1}, but now as a function of number of CRs injected per turbulent turnover time, $n_{\rm CR}/\tau$. Note that the parameters shown here for the low $\Mao$ trials are slightly different than in \autoref{fig:time_c1}. We find that the fitted diffusion coefficients in both parallel and perpendicular directions have little to no variation across our range of $n_{\rm{CR}}$.}
    \label{fig:part_c}
\end{figure*}

Our next convergence test is with regard to the rate at which we inject CR packets from each source, and thus the temporal resolution with which we sample the structure in our \textsc{criptic} simulations. We parameterise this quantity in terms of the number of packets $n_{\rm CR}$ injected per turbulent turnover time $\tau$; we test values of $10^4 - 10^7$ in steps of 10. \autoref{fig:part_c} shows the results of the particle resolution testing. We see that there is little variation to the extracted diffusion coefficients regardless of the particle rate. We use a particle rate of $10^6 \ n_{\rm{CR}} / \tau$. It should be noted here that the injection rate only controls the number of sample particles injected at a pre-set source location, so increasing $n_{\rm{CR}}$ effectively increases the number of CR packets on a single fieldline. Thus it is perhaps not surprising that the results are not sensitive to this choice, as long as $n_{\rm CR} \gg 1$.

\subsection{Source resolution} 

\begin{figure*}
    \includegraphics[width=0.95\textwidth]{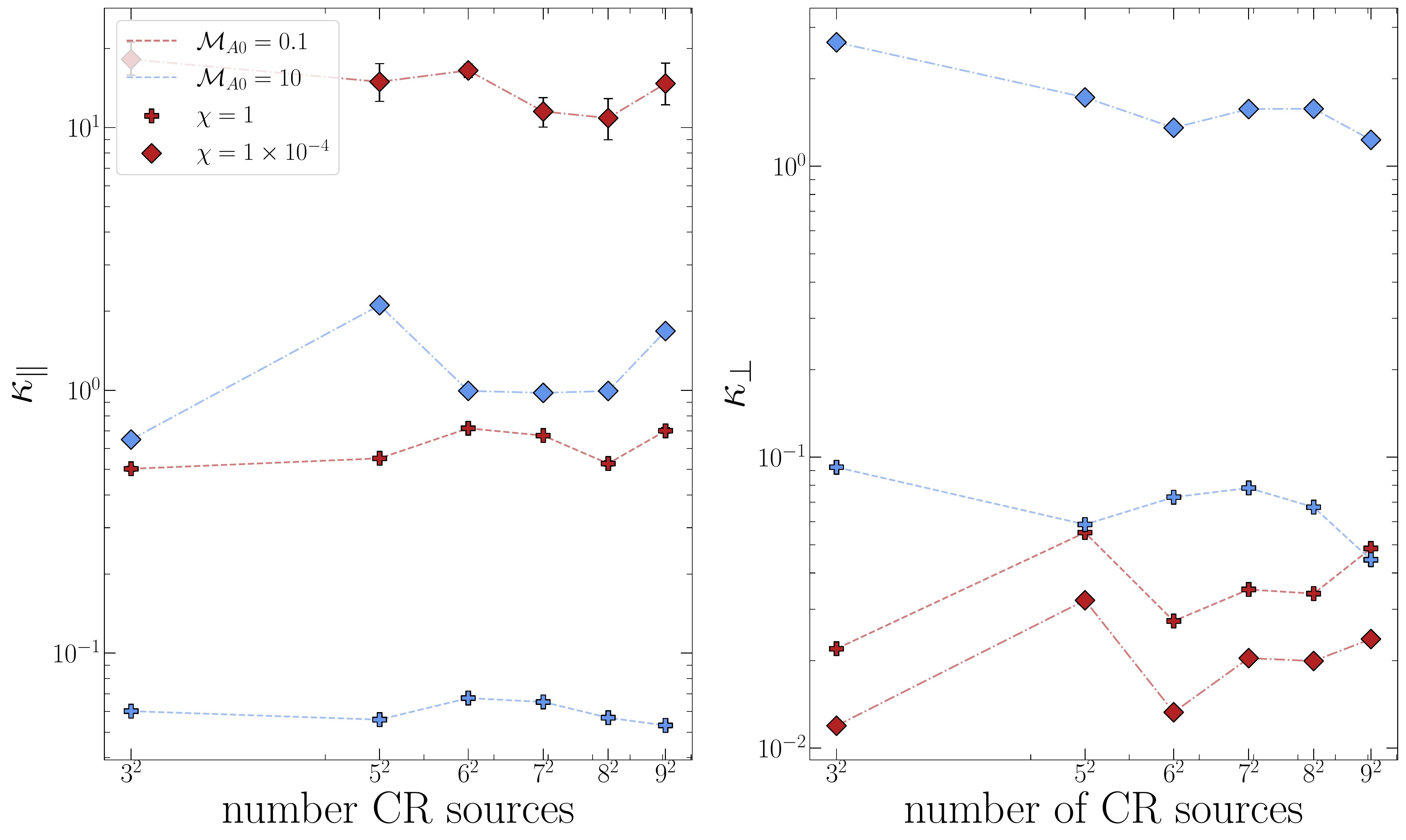}
    \caption{Similar to \autoref{fig:part_c} except the x-axis shows the number of CR source locations which are all configured in a square grid through the $z = - L$ plane (the base of our simulation box). We see good convergence above $7^2$ sources.}
    \label{fig:source_c}
\end{figure*}

Our final test evaluates convergence of the diffusion coefficients in the number of CR injection sites, which is a measure of how well we sample the turbulence spatially (or how our injection configuration changes the nature of the measured diffusion). The geometry of the CR sources is in all cases a uniformly-spaced square grid placed at the lower boundary ($z = -L$) of the simulation box; we vary the number of sources in this grid from $3^2$ to $9^2$.
\autoref{fig:source_c} shows the results of this test, which show reasonable convergence beyond $7^2$ sources. We use a $9 \times 9$ grid of CR sources for all trials in the main text.

%%%%%%%%%%%%%%%%%%%%%%%%%%%%%%%%%%%%%%%%%%%%%%%%%%

% Don't change these lines
\bsp	% typesetting comment
\label{lastpage}
\end{document}